\documentclass[12pt]{article}

\usepackage{epsfig}
\usepackage{cite}
\usepackage[normalem]{ulem}
\usepackage{amsmath, amssymb, amsfonts}
\usepackage{cancel}
\usepackage{color}
\usepackage{latexsym}
\usepackage{graphicx}
\usepackage[colorlinks,bookmarks]{hyperref}
\hypersetup{pdfpagemode=UseNone, pdfstartview=FitH, linkcolor=blue,
            citecolor=red, urlcolor=blue}

\def\be{\begin{equation}}
\def\ee{\end{equation}}
\def\ba{\begin{eqnarray}}
\def\ea{\end{eqnarray}}

\def\bdm{\begin{displaymath}}
\def\edm{\end{displaymath}}
\def\la{~\mbox{\raisebox{-.6ex}{$\stackrel{<}{\sim}$}}~}
\def\ga{~\mbox{\raisebox{-.6ex}{$\stackrel{>}{\sim}$}}~}
\def\bq{\begin{quote}}
\def\eq{\end{quote}}

 at 10truept

\newcommand{\rmd}{\mathrm{d}}

\newcommand{\comm}[2]{\left[ #1 , #2 \right]}
\renewcommand{\[}{\left[}
\renewcommand{\]}{\right]}
\renewcommand{\(}{\left(}
\renewcommand{\)}{\right)}
\newcommand{\vk}{\vec{k}}

\newcommand{\vphi}{\varphi}

\newcommand{\bea}{\begin{eqnarray}}
\newcommand{\eea}{\end{eqnarray}}

\newcommand{\bi}{\begin{itemize}}
\newcommand{\ei}{\end{itemize}}

\newcommand{\beq}{\begin{equation}}
\newcommand{\eeq}{\end{equation}}
\newcommand{\beqa}{\begin{eqnarray}}
\newcommand{\eeqa}{\end{eqnarray}}

\def\la{~\mbox{\raisebox{-.6ex}{$\stackrel{<}{\sim}$}}~}
\def\ga{~\mbox{\raisebox{-.6ex}{$\stackrel{>}{\sim}$}}~}
\def\d{{\rm d}}

\newcommand{\vx}{\vec{x}}

\newcommand{\vq}{\vec{q}}

\newcommand{\calH}{\mathcal{H}}

\def\ltap{\ \raise.3ex\hbox{$<$\kern-.75em\lower1ex\hbox{$\sim$}}\ }
\def\gtap{\ \raise.3ex\hbox{$>$\kern-.75em\lower1ex\hbox{$\sim$}}\ }
\def\gl{\ \raise.5ex\hbox{$>$}\kern-.8em\lower.5ex\hbox{$<$}\ }
\def\roughly#1{\raise.3ex\hbox{$#1$\kern-.75em\lower1ex\hbox{$\sim$}}}

\begin{document}

\thispagestyle{empty}
\begin{flushright}
November 2020 \\
\end{flushright}
\vspace*{.25cm}
\begin{center}
  
{\Large \bf Rollercoaster Cosmology}

\vspace*{1cm} {\large Guido D'Amico$^{a, }$\footnote{\tt
damico.guido@gmail.com} and Nemanja Kaloper$^{b, }$\footnote{\tt
kaloper@physics.ucdavis.edu} }\\
\vspace{.3cm} {\em $^a$Dipartimento di SMFI dell' Università di Parma and INFN, Gruppo Collegato di Parma, Italy}\\
\vspace{.3cm}
{\em $^b$Department of Physics, University of
California, Davis, CA 95616, USA}\\

\vspace{1.5cm} ABSTRACT
\end{center}
Does inflation have to happen all in one go? The answer is a resounding no! All cosmological problems
can be solved by a sequence of short bursts of cosmic acceleration, interrupted by short epochs of decelerated expansion. The spectrum of perturbations will still match the CMB and LSS if the earliest stage of the last ${\cal O}(50)-{\cal O}(60)$ efolds is at least ${\cal O}(15)$ efolds long. Other stages can be considerably shorter. But as long as they add up to ${\cal O}(50)-{\cal O}(60)$ efolds and the stages of decelerated expansion in between them are shorter and also overall last less, the ensuing cosmology will pass muster. The presence of the interruptions resets the efold clock 
of each accelerating stage, and changes its value at the CMB pivot point. This change opens up the theory space, loosening the bounds. In particular some models that seem excluded at ${\cal N}=60$ fit very well
as shorter stages with ${\cal N}=30$. Interesting predictions are that both the scalar and tensor spectra of perturbations are rapidly modified at short 
wavelengths. In the simplest cases the perturbations are suppressed 
relative to the perturbations at large scales just because when they freeze out, the background curvature is smaller. The modes which do leave the horizon however
will remain frozen as long as the subsequent intervening stages of decelerated expansion remain short. These 
features could be tested with future CMB spectroscopy searches and with short wavelength primordial gravity probes. The spatial curvature in these models can be larger than the largest wavelength scalar perturbations, because
$\Omega_{\tt k}$ evolves differently than the scalar perturbations $\frac{\delta \rho}{\rho}|_{\tt S}$. Finally, with many short stages of accelerated expansion, the abundance of reheating products from previous accelerated stages does not get completely wiped out. This implies that
the universe may contain additional populations of particles, more rare than the visible ones, or even primordial
black holes, created during a late decelerated epoch before last reheating, which 
may be dark matter.

\vfill \setcounter{page}{0} \setcounter{footnote}{0}

\vspace{1cm}
\newpage

\hfill \parbox{12.6cm}{
{\it An astronomer, a physicist and a mathematician are riding a train}
{\it through Scotland.}\\
{\it The astronomer looks out the window, sees a black sheep, and exclaims,} \\ 
{\it ``Hey! They've got black sheep in Scotland!"}\\
{\it The physicist looks out the window and corrects the astronomer,}\\
{\it  ``Strictly speaking, all we know is that there's at least one black sheep}\\
{\it in Scotland."}\\ 
{\it The mathematician looks out the window and corrects the other two,}\\
{\it ``Strictly speaking, all we know is that at least one side of one sheep}\\
{\it \, is black in Scotland."}\\
}
\begin{flushright}
\hfil{\it (A joke from the interwebs.)}
\end{flushright}

\section{Introduction}

Flatness, homogeneity and isotropy of the  universe on  the largest scales strongly support the idea that
the universe was shaped by inflation: a period of exponential expansion with a nearly constant Hubble parameter
which lasted between 50 and 60 efolds \cite{Guth:1980zm, Linde:1981mu,Albrecht:1982wi}. The usual assumption is that the epoch of inflation is uninterrupted, all of it
happening in one go. The idea is that the universe is dominated by an agent which behaves like vacuum
energy, with nearly constant energy density, whose relativistic pressure acts as antigravity and blows the space
up \cite{Guth:1980zm, Linde:1981mu,Albrecht:1982wi}. Realizing such vacuum energy models which are metastable, but very long lived, is quite challenging. While 
there are many proposals, most run into troubles with  quantum field theory (QFT). The issue is typically with 
the requisite longevity of the inflationary stage; building shorter-lasting models is generally easier.

Here we argue that a different approach may be possible. Instead of a single period of long inflation the usual cosmological problems are solved by a sequence of short bursts of accelerated expansion, interrupted by interim short stages of different cosmological evolution. Indeed we imagine that after each brief period
driven by a field on a shallow slope, a stage of accelerated expansion ends and is followed by a stage of deceleration (where
we take the decelerating stage to be radiation-dominated for simplicity\footnote{Other agents may be dominating evolution
during interim deceleration stages, such as matter, or a `gas' of extended objects and so on. 
They would modify some of the
details of the model predictions such as the spectrum of perturbations at relevant scales. However we leave the study of those interesting details for a later time.}. Radiation is generated by the decay of the inflating agent, and dominates but briefly, until the next stage of cosmic acceleration bursts again\footnote{It may also be possible that some stages of deceleration are very long, delaying a stage of accelerated expansion for long enough for galaxies and stars to form and life to rise. Such as in our universe...}. 
And so on, pump-and-dump. We dub this dynamics the rollercoaster cosmology.

This cyclic repetition of acceleration and deceleration of cosmic expansion behaves very similarly to uninterrupted 
inflation when it comes to the background. The many short epochs of cosmic acceleration add up to dilute initial anisotropies, inhomogeneities
and curvature, each a bit at a time, and the intervening radiation epochs do not support their regrowth in between the subsequent bursts of accelerated expansion.
Therefore the  repetition of the accelerating stages adds up to the total dilution of the initial conditions with an effect very similar to a single stage of long inflation.
We show explicitly that as long as the total number of inflationary efolds summed over many stages of our rollercoaster universe adds to $\ga 50$, it  suffices to 
smoothen and flatten the universe as required. Basically, what 
matters is the integrated effect of accelerating stages. As long as enough total expansion occurs under acceleration, and the total duration of the decelerating interruptions is not too long, the background flattens out.

However inflation doesn't only erase generic initial conditions. It also sets up the stage to generate the seeds for
subsequent structure formation in the late universe. The quantum fluctuations of the agent which causes the rapid expansion originate in near-Bunch-Davies (BD) quantum vacuum \cite{Bunch:1978yq} due to the Heisenberg uncertainty principle. They imprint curvature perturbations that generate weak gravitational wells in the late universe to initiate 
formation of galaxies, and at longer scales, the observed CMB anisotropies \cite{Mukhanov:1981xt}. The rollercoaster cosmology should do
at least as well to be a candidate for a realistic early universe cosmology. We develop a simplified approach to
this problem and show that this indeed happens.
An initial shorter burst of early cosmic acceleration sets 
up a BD vacuum \cite{Kaloper:2018zgi} where the fluctuations originate, and propagate through the many subsequent
stages of evolution.

The long wavelength perturbations behave almost exactly the same as the long wavelength 
perturbations in inflation. After they emerge from the quantum vacuum they 
undergo wavelength stretching induced by the expanding background. Once their wavelength becomes
larger than the Hubble length of the background, they freeze out. After the end of each burst of cosmic acceleration, during 
the intervening short epochs of radiation, these modes remain frozen because their wavelength remains longer
than the background curvature. All that happens to them is a phase shift, which however does not affect the amplitude. Thus as long as the total duration of all short bursts of cosmic acceleration is $\ga 50$ efolds, these modes
behave just like in inflation\footnote{In fact, for very long-wavelength modes, fewer than $50$ efolds may suffice.}.

The short wavelength modes however are very different. Below a critical wavelength, set by the duration of the 
individual bursts of cosmic acceleration, these modes never leave the horizon. Thus they never freeze out, and their amplitude is
inflated away. Hence in the simplest realization of rollercoaster cosmology, with interim radiation eras, there is a sharp cutoff in the spectrum of perturbations $P(k)$, where the power in modes above 
$k_{critical}$ quickly diminishes with increasing $k$, and remains approximately constant -- i.e. scale invariant -- for $k < k_{critical}$. This is true for both scalar and tensor modes, and provides observational means to test the proposal against 
standard inflationary paradigm. Alternatively this adds a mean to test inflation observationally: in the simplest realizations of our proposal,
the spectra of perturbations will have very little power at very short distances, whereas in the case of a single long stage of inflation, with scale invariant spectrum, the short wavelength power will remain roughly the same as at large wavelengths in both scalar
and tensor modes. This could be directly tested by short wavelength searches for primordial perturbations, 
specifically the future CMB spectroscopy searches \cite{spectr}, and the short wavelength primordial gravity wave probes in experiments \cite{Cornish:2002fg,Brazier:2019mmu,Arzoumanian:2020vkk}. 

The modes with $k$ in an intermediate range however may freeze out and thaw multiple times 
during the evolution due to the sequence of acceleration and deceleration during which $k$ could be crossing in and out of the horizon.
Clearly, in our case, the power in these modes would diminish relative to the long wavelength modes, and hence
these modes must not occur at scales corresponding to CMB anisotropies. 

As we already noted however, one could imagine different scenarios, where the decelerating stages 
would not be radiation-dominated. In such cases the spectra at short wavelengths might even be amplified relative to long inflation.
The conclusion remains - those are not the modes we want to occur at CMB scales. If the stage of accelerated expansion is sufficiently long, $\ga {\cal O}(15)$ efolds, and the subsequent radiation era short, $\la {\rm few}$, the modes 
which correspond to CMB scales will generically be safe: they will remain frozen. 
The important point is, however, that the overall spectra beyond CMB would differ from those in long inflation. Hence we will settle for a simplest demonstration
of the spectral features here, using short radiation eras to interrupt acceleration for simplicity, and leave more general cases for future study.

We will also see that due to the presence of the  intervening short radiation epochs in between the bursts of accelerated expansion, the correlation between the evolution of background cosmic curvature 
$\Omega_{\tt k}$ and the largest wavelength scalar perturbations is disrupted. While the perturbations with superhorizon wavelengths
retain constant amplitude during those radiation eras, the curvature term $\Omega_{\tt k}$ ``thaws" and grows. 
Thus rollercoaster evolution, while making $\Omega_{\tt k}$ small, will suppress it more slowly than long smooth
inflation. So it is possible to have
$\Omega_{\tt k}$ considerably different than $\frac{\delta \rho}{\rho}|_{\tt S}$. In  a single stage long inflation, instead, 
these quantities are expected to be correlated \cite{Knox:2005hx,Kleban:2012ph,Efstathiou:2020wem}. 

Another important implication of the interruptions is that the fitting of the model predictions has to be done slightly differently than 
in long smooth inflation even for modes which are frozen during eras of acceleration, and remain frozen until after the very late times. 
This is necessitated by the fact that the total amount of efolds is a `composite' of many different epochs, each with its own internal efold clock
and internal dynamics. Thus the events which determine observables do not simply occur at ${\cal N}$ efolds before the end of inflation. The evolution is more like a symphony with multiple movements, and a clock reset at the beginning of every new stage of acceleration, where the show restarts anew. Thus the dynamical clock  
restarts and counts ${\cal N}$ efolds before the end of any particular stage. The interruptions reset the
clock of each accelerating stage, due to the locality of the potential in field space, and therefore locality in time. 
Specifically this changes the value of ${\cal N}$ at the CMB pivot
point, where the CMB observations are fitted to the theory. This can significantly open up the theory space, loosening the bounds. 
In particular some models that seem excluded at ${\cal N} = 60$ fit very well as shorter stages with ${\cal N} = 30$. Furthermore
this also demonstrates, for example, that an oft-heard comment these days, that data favor 
the Starobinsky model \cite{Starobinsky:1980te} (which
the data certainly agree with well), has a hidden assumption: that inflation was long and smooth, and that the CMB corresponds
to 50 -- 60 efolds before inflation ended. As we see, this is not (yet) necessary. 

In this paper we will present a very simple mechanism illustrating these ideas. We will not undertake a detailed QFT construction  
of a microscopic theory that may be responsible for such dynamics. Our main goal here is to set up the kinematics 
of the rollercoaster cosmology in order to illustrate that it can be an interesting candidate for the description 
of the early universe.  We will leave more precise model building for future work. 

First we will show that a sequence of short bursts of accelerated expansion interrupted with short interim epochs of radiation does suffice to solve the cosmological problems with initial conditions as long as the
total expansion of the universe during acceleration epochs  adds up to $\ga 50$ efolds, and the radiation stages last less. 
We will use the horizon
and the curvature problems as the two cases in point. 

Secondly we will study perturbation evolution 
by quantizing a light scalar field and finding the power spectra of its quantum fluctuations in  the BD vacuum.
For simplicity, in this work we will assume that the epochs of acceleration and deceleration repeat regularly
at the same intervals in conformal time, yielding a regular structure of potential barriers and wells 
through which the perturbations propagate in conformal time. We will then compute power of the fluctuations -- i.e. amplitude squared -- 
as a function of comoving momentum. We imagine that during each accelerating stage the curvature is fed
by a dominant inflaton during that time, which quickly  decays away, and is replaced by a new one at the next stage.
We will not engage in a detailed ``roll call" of the inflaton, or inflatons, required to drive the sequence of accelerating stages, merely assuming that a multifield framework may be constructed.
We will ignore most of the `edge' and mixing effects, which may be important, again solely for the reasons of simplicity.
Probably the most important of these effects is the generation of isocurvature perturbations, generic in multifield inflation models. Since the isocurvature perturbations are strongly constrained, this could lead
to important limits on model building. Yet it is clear this does not yield an insurmountable problem since
the isocurvature perturbations are generated predominantly at turns in the field space, and as long as the trajectories
are piecewise ``straight" for long enough, the isocurvature perturbations will be suppressed in the observed regime,
completely disappearing in the single field limit \cite{maartens}.
In the case of tensor perturbations, the scalar field is simply the intensity of the primordial gravity waves. After studying perturbations, we will also sketch out
a possible realization of rollercoaster stages by some simple field theory models such as flat power laws and exponentials, 
mainly to illustrate the difference of the observables with those from long smooth inflation.

As we noted, we will also show that $\Omega_{\tt k}$ may be larger than $\frac{\delta \rho}{\rho}|_{\tt S}$. This may be 
useful in searching for the sources of small anomalies in the data (see, e.g. \cite{silky}).
We will also mention the possibility that 
because the evolution of the universe involves many short stages of accelerated expansion, unlike in long inflation the 
abundance of reheating products from previous accelerated stages need not get completely wiped out. Hence 
the universe may contain additional populations of particles, more rare than the visible ones, which if heavier may be dark matter. This dark
matter might even be quite exotic, as some examples discussed recently in the review \cite{watfries}. 

We must mention here very important precursor work which we found after realizing that rollercoaster
cosmology is a plausible scenario of cosmic origins. Exercising due diligence we found papers featuring aspects of 
rollercoaster cosmology, some going 
as far back as the time of the inception of inflationary cosmology, where the idea of double inflation
was discussed \cite{andrei,stary,kofman,silkyturn,sabob,slava1,polstar}, where two stages of inflation are separated by matter domination. The 
effects of such an interruption of inflation was studied in \cite{punct,misao}. Most notably, dynamics of multistage inflation
was pointed out in \cite{damsla,grahshar,clifton,westy}.  
In \cite{grahshar}, QFTs supporting connected multiple bursts of shorter stages inflation were discussed. Overall, this scenario involved inflation with a variable spectral 
index due to occasional stages of faster rolling of the inflaton, with the spectrum of perturbations following closely that of inflation, except for 
the localized breaks in $n_S$.  This approach does not explicitly consider a more radical possibility, which is that inflation ends and
restarts, leading to the rollercoaster evolution. However \cite{clifton,westy} 
note that non-accelerating stages separating epochs of short inflation may 
have occurred while still fitting the observations. Here we carry out an extended analysis of the full features of rollercoaster cosmology showing how the background problems are
resolved and the observations consistently reproduced, and giving a 
discussion of new predictions at short scales, significantly expanding on the work published to date. 
We also note the importance of stage-to-stage efold clock reset for matching data to theoretical models,
which is done differently than in long smooth inflation, with significant implications for model building.

We start with setting up the exact background geometry in the 
limit where the accelerating phases are de Sitter, and carry out the evaluation of the quantum fluctuations and their imprint onto the
background curvature using the gauge invariant variables, setting the stage for the computation of the cosmological observables.
We discuss implications for model building and for phenomenology and cosmology, but will leave a detailed construction of
a candidate rollercoaster theory for future work. Curiously, much of the followup to the important early papers \cite{andrei,stary,kofman,silkyturn,sabob,slava1,polstar,damsla,grahshar}
seems to have focused on determining how some -- 
typically small -- disruptions early in inflation can generate effects that might be used to fit alleged
anomalies in the sky. This application is valid, but we feel focusing on it alone risks missing the forest for a tree. Spectral disruptions
might lie in wait at other than the largest scales in the sky, and be a possible handle to discover much more interesting early universe 
dynamics.

\section{Horizon and Flatness Problems} 

For a classic review 
of the cosmological problems, which inflation was designed to solve, the reader may consult \cite{Linde:2005ht}. Here we will focus on the horizon and flatness 
problems, and explain how they are solved in rollercoaster cosmology. Other problems discussed in 
\cite{Linde:2005ht} can be solved similarly. 

The horizon problem concerns the explanation of the size of the causally connected domain of the universe. Observations show that the 
universe is correlated over scales at least as large as the current Hubble scale. Using the FRW metric, and ignoring the
spatial curvature, this means that the correlations are seen at scales $\ell \sim 1/H_{now}$ where $H_{now}$ is the Hubble parameter
now. If we extrapolate this scale backward in time, using the standard FRW scaling $\ell \sim \frac{1}{H_{now}} \frac{a(t)}{a_{now}}$ where
$a(t)$ is the scale factor at a time $t$ and $a_{now}$ its value now, and compare it to the size of the causal domain given
by the particle horizon
 \be 
 L_H = a(t)\int_{t_1}^t\frac{\d t^\prime}{a(t^\prime)} \, ,
 \label{horizon}
 \ee
 we see that in universes dominated by normal matter\footnote{Radiation, dust, $\ldots$}, $\ell \sim t^{2/3(1+w)} \sim (1/H)^{2/3(1+w)}$ and $L_H \sim t \sim 1/H$, where $w \ge 0$ is the equation
 of state parameter for normal matter. Hence 
 \be
 \ell/L_H \sim H^{(3w+1)/(3w+3)} \, ,
 \ee
 which means that $\ell$ grows more slowly
 than the horizon when $w > -1/3$. So for the ratio to be unity now, in a universe filled with normal matter the initial correlation size $\ell_{in}$ must be 
 many orders of magnitude larger than the initial causal domain size $L_{Hin}$. In other words, in a universe filled with normal
 matter the initial correlations must be finely tuned over a region much larger than the causal domain. 
 
As in inflation, here we take the point of view that normal matter cannot dominate the universe throughout its past evolution.
Instead more exotic materials must exist, which accelerate the evolution of $a(t)$ allowing a small initial correlated domain to keep up with
the horizon, and maintaining the correlations over at least the horizon-sized regions. As we noted, we imagine that the agent which
facilitates this is like the inflaton, but more unstable, decaying faster. However there are many consecutive epochs. 
So let us consider a rollercoaster ride of such bursts driving the universe from radiation to accelerated expansion to radiation to accelerated expansion and so on. To see how the particle horizon evolves, it suffices to note that in an expanding universe, the integral in (\ref{horizon}) is dominated by the lower limit. 
Evaluating it in the first segment of a multiple stage rollercoaster, where each segment consists of a radiation phase followed by an accelerated expansion phase, we find that
 \be
\int_{t_0}^t\frac{\d t^\prime}{a(t^\prime)} \simeq \frac{1}{\sqrt{H_1 H}} \, ,  
\ee
where $H$ is the Hubble parameter at the beginning of the first radiation stage, and $H_1 \la H$ its value at the onset of the first subsequent 
stage of accelerated expansion $t_1 > t_0$. The overall scale factor is just a product of the scale factor variations at each individual stage, $a(t) = \Pi_{j} \frac{a(t_{j+1})}{a(t_j)}$. 
Thus the particle horizon over many rollercoaster stages evolves simply as $L_H \sim a(t)/\sqrt{HH_1} \la a(t)/H_1$. Hence, 
\be
\ell/L_H \ga \ell_{in} H_1 \, .
\ee
To ensure that the universe is causally connected we only need to pick the initial correlation length to be given by the inverse 
of the Hubble parameter at the beginning of the first burst of accelerated expansion. This choice solves the horizon problem. 

Consider now the flatness problem. The issue now is that the contribution of the spatial curvature to the critical energy density today is 
bounded by $1/100$, 
\be
\Omega_{{\tt k}~now} = \frac{\tt k}{a_{now}^2 H_{now}^2} \la 0.01 \, .
\ee
We realize this is a problem when we compare $\Omega_{{\tt k}~now}$  to its value at a previous epoch $\Omega_{\tt k}(t_*) = \frac{{\tt k}}{a^2_* H^2_*}$. Their
ratio is $\Omega_{{\tt k}~now}/\Omega_{{\tt k}*} = \frac{ a^2_* H_*^2}{a_{now}^2 H_{now}^2}$, which in a universe with normal matter with the equation of state parameter 
$w \ge 0$ is 
\be
\frac{\Omega_{{\tt k}~now}}{\Omega_{{\tt k}*} }=  \Bigl( \frac{ H_*}{H_{now}}\Bigr)^{2(1+3w)/3(1+w)} \, .
\ee
For $w \ge 0$ this means that at earlier times when $H_*$ is large, $\Omega_{{\tt k}~now} \gg \Omega_{{\tt k}*}$. So for $\Omega_{{\tt k}~now}$ to be less than $0.01$ today, its
value at previous epochs had to be miniscule. This problem is solved in rollercoaster cosmology by the repeated short stages of accelerated expansion, because early on 
the scale factor includes exponential factors from each burst of accelerated expansion. Indeed, if we take an elementary rollercoaster segment of a short radiation stage
followed by a  short stage of accelerated expansion, the scale factor evolves according to $a_j = a_{j-1} e^{N_j} = a_{j-2} \sqrt{H_{j-2}/H_j}$, where $N_j$ is the number of efolds 
during the burst of accelerated expansion, $H_j$ the Hubble parameter during the $j^{th}$ burst, and $H_{j-2}$ the Hubble parameter during the previous burst, and  
$\sqrt{H_{j-2}/H_{j}}$ is the scale factor growth during the intervening radiation era. 

Multiplying these for a sequence of segments we find that at the end
of the rollercoaster phase of $n$ segments the product $a_n H_n$ has grown by
\be
a_n H_n = a_1 H_1 \sqrt{\frac{H_n}{H_1}} e^N \, ,
\label{kinfl}
\ee
where $N$ is the cumulative number of efolds due to all the periods of accelerated expansion, $\sqrt{H_n/H_1}$ the correction during the intervening radiation epochs
and $a_1 H_1$ the starting value at the first burst of accelerated expansion. Thus at the onset of the cosmology dominated by  normal matter, $\Omega_{{\tt k}*}$ is down by a factor of $\frac{H_n}{H_1} \, e^{2N}$.
Note that during the single long stage of inflation, this factor would be larger, $\sim e^{2N}$. The reason is that while $\Omega_{\tt k}$
is diluted during inflation, it grows during radiation era because the curvature term ${\tt k}/a^2$ dilutes only as $1/a^2$, as opposed to $1/a^4$ for radiation. 
Hence their ratio, ${\tt k}/a^2 \rho_{radiation} \sim \Omega_{\tt k}$ grows as $a^2 \sim 1/H$. This means that while the curvature problem is solved,
for a fixed number of efolds the curvature $\Omega_{\tt k}$ may end up enhanced relative to its value during a  single long stage of inflation 
of the same total duration. Put another way, in the rollercoaster cosmology, the curvature problem is a stronger 
problem that the horizon problem: once it is solved, the horizon problem is over-solved. We 
will discuss below how this removes degeneracy between $\Omega_{\tt k}$ and $\frac{\delta \rho}{\rho}|_{\tt S}$. 

We note here that the discussion about the evolution of the curvature contribution to the cosmic expansion, $\Omega_{{\tt k}*}$, might 
trigger an alarm that in fact the rollercoaster evolution might be very constrained from the get go. Indeed, from Eq. (\ref{kinfl}) 
it seems that if the ratio $H_n/H_1$ is too small, the resulting curvature $\Omega_{{\tt k}~now}$ might end up being too big. 
This insinuates a bound on $H_n/H_1$, which could constrain rollercoaster severely, since during each decelerating stage
$H$ changes as inverse two powers of the scale factor. However, the bound only arises if we imagine that 
${\Omega_{{\tt k}*} } \simeq 1$ at the onset of rollercoaster evolution. There is no reason for this to be the case: the rollercoaster 
phase may have been preceded by a longer stage of inflation, which would make the curvature term at the 
onset of the rollercoaster safely small. In fact, even ${\cal O}(10)$ efolds may do fine, reducing 
${\Omega_{{\tt k}*} }$ by $10^9$. Since ${\cal O}(10)$ efolds is likely needed to single out the Bunch-Davis state as the
approximate initial quantum state of the rollercoaster universe \cite{Kaloper:2018zgi},  this means that the bounds on subsequent 
rollercoaster evolution would be automatically weak. Nevertheless, the intrinsic growth of $\Omega_{\tt k}$ during intervening radiation
eras makes $\Omega_{\tt k}$ intrinsically historically sensitive. It is possible that due to slowed down suppression rate, the marginal cases
of largeish $\Omega_{\tt k}$ come out of the rollercoaster evolution. Since such primordial values may not be negligibly small, 
the net observed value of $\Omega_{\tt k}$ might exceed the intrinsic value set by the density contrast 
$\frac{\delta \rho}{\rho}|_{\tt S} \sim 10^{-5}$ at horizon scales. This may be a useful test to distinguish rollercoaster 
from a single stage long inflation \cite{Knox:2005hx,Kleban:2012ph,Efstathiou:2020wem}.

\section{Rollercoaster Architecture}

The next step is to show that the quantum curvature fluctuations still provide the mechanism for the generation of the
seeds for galaxy formation, and hence also source the anisotropies in the CMB, in the regime \underbar{\it where we observe them}.
This is the important point which we cannot stress enough. A key support for the inflationary paradigm is that it produces a nearly scale invariant
spectrum of perturbations of the underlying geometry \cite{harrison,zeldovich} which produces a very minimal local and causal 
explanation of the cosmological observations at the largest scales, which is under calculational control in perturbative EFT\footnote{There are, 
however, theoretical challenges in setting up UV-complete QFTs coupled to gravity that support such EFTs on long-lasting inflationary  
backgrounds. It seems not everything goes.}. As is well known, the idea is that the inflationary expansion irons out the initial, 
possibly very large, inhomogeneities and anisotropies, and replaces them by the smaller, almost scale invariant fluctuations generated
by the quantum uncertainty principle of the inflaton in the BD vacuum. These feed into the curvature, and are frozen by the rapid expansion of the background which decoheres the fluctuations and conserves their amplitudes. 

However -- as we stressed -- in principle it 
suffices if this only happens in the regime of scales which we have observed so far. If there are deviations and distortions of the spectrum at
shorter scales, not only is that perfectly fine, but in fact may be an exciting new avenue for testing cosmogenesis in the future. 
Indeed, interrupting inflation is not detrimental to it, but may help induce small deviations which could help with the 
alleged anomalies in the data \cite{punct}. Here however we will take the extreme approach and 
explore the maximal transient deviations from inflation, allowing not 
only for interruptions, but dramatic breakdowns of inflationary expansion, and yet show that this nevertheless does allow for spectra which
fit the data \underline{\it observed to date}. 

In order to accomplish this, we need to set up the stage for the study of perturbations. We start with constructing the background solutions. 
Since cosmological expansion is conformal to a time translation, and since conformally flat metrics are a simple background to study
perturbations, we will work with conformally flat FRW metrics. This simply means, as in inflation, that we assume that prior to the epoch during
which the fluctuations were created, the rollercoaster evolution already took place for some time, erasing the initial
data and flattening the universe just enough to ignore the initial curvature, anisotropies and inhomogeneities. A single stage of inflationary expansion
of ${\cal O}(10)$ efolds should suffice \cite{Kaloper:2018zgi}. This stage also prepares the initial quantum state to be the BD vacuum. 

Subsequently, we will imagine that the evolution comprises a sequence of 
inflating stages separated by stages of decelerating expansion, just like a rollercoaster ride at a boardwalk carnival. 
Since this happens at high energies, we will 
simplify the description by imagining that the decelerating stages are radiation dominated, by the radiation generated instantaneously at the
transition from the preceding inflating stage.We stress that our use  of radiation is for simplicity reasons.
One could imagine different agents dominating intermediate 
decelerating stages, such as matter. Matter eras would arise automatically if the reheating at each stage of acceleration is inefficient and the post-acceleration geometry is  dominated by the former inflaton oscillating 
around its minimum. During such stages perturbations might grow a little, leading to interesting signatures. We will not study the details of such dynamics here, but note that some studies are given in \cite{stary,slava1,polstar}. We
do believe further work on this subject is warranted in order to determine the observational signatures of our proposal in full generality.

Further, we will approximate each inflating stage by a section of exactly de Sitter background.
A simple realization of a rollercoaster model is to imagine such quasi-de-Sitter stages to arise as a leading order approximation of the dynamics 
where the universe is controlled by a spectrum of inflatons, with masses which are 
parametrically different from each other. The earlier stages are controlled by the heaviest inflatons, which roll slowly until $m<H$, and then
decay very quickly into radiation. The lighter inflatons also roll, but far too slowly to do anything during the epoch controlled by the heavier
one, and with too small potential energy. These remain safely decoupled until later, behaving much like a small residual cosmological constant. 
And so on -- the inflatons taking turns at controlling the evolution, each being poised to dominate
the energy density a short while after the decay of the previous one in the sequence. 

One may worry that the many ultralight inflaton-to-be's, although
having subleading zero modes during a stage a field of mass $m$ dominates, may have large fluctuations that mix, and distort, the adiabatic fluctuations of the dominant field.
However, since they only imprint on the curvature to linear order via  mixing terms induced by their time derivatives, since they are very deep in slow roll their contributions to isocurvature perturbations will remain suppressed.
Indeed, it has been shown \cite{maartens} that in multifield models isocurvature perturbations are generated, 
but not when the inflaton follows straight paths in the field space. This means, that the isocurvature perturbations may be significant,
but by our assumptions not at the currently accessible scales, as long as the CMB and LSS are generated during a rollercoaster
epoch that lasted at least 15 efolds. In what follows, we will ignore isocurvature perturbations for simplicity, even though we believe 
that a detailed and precise analysis is warranted to obtain predictions for future searches.

With these assumptions, we can readily construct the background geometry using Israel junction conditions which teach us
how to sew the different phases together \cite{Deruelle:1995kd}. Since we are assuming for simplicity that evolution is adiabatic,
so that the energy density is continuous across the junctions, both $a$ and $a'$ are continuous -- regardless of whether we use
comoving or conformal time. The only jump happens in the equation of state -- i.e. in $a''$, and it is bounded. 
This means that we can use the metric ans\"atz 
\be
ds^2 = a^2(\tau) [-d\tau^2 + d\vec x^2] \, .
\label{met}
\ee
The task is now to determine the scale factor $a(\tau)$. 

In each stage of accelerated expansion, the conformally flat FRW scale factor is $a \sim e^{Ht}$, which means 
that after the transition to conformal coordinates it is $a = \frac{1}{H(\hat \tau_i - \tau)}$. Exponential accelerated expansion in these coordinates is the 
limit $\tau \rightarrow \hat \tau_i - \epsilon$. On the other hand, in the radiation stages $a \sim \sqrt{t}$, which translates to 
$a = a_0 (\tau + \hat \tau_r)/2$.
Since $a_0$ is a gauge parameter, we will fix it to be $a_0 = 2/H$. The integration 
constants at each phase $\hat \tau_i, \hat \tau_r$ are determined sequentially, using the junction conditions, according to
\be
a_n(\tau_n) = a_{n+1}(\tau_n) \, , ~~~~~~~~~~~ a'_n(\tau_n) = a'_{n+1}(\tau_n) \, ,
\label{bcs}
\ee
where $\tau_n$ is the instant where the $n^{th}$ and the $(n+1)^{st}$ stage are conjoined. The integration constants $\hat \tau_i, \hat \tau_r$ are determined
by the clock synchronization between adjacent stages. 

To determine the full metric, rewrite (\ref{met}) as 
\be
ds^2 = \hat a^2(\tau) \frac{[-d\tau^2 + d\vec x^2]}{H^2}\, .
\label{metric}
\ee
Here $H$ is the Hubble parameter in the pre-inflationary stage. 
Note that the coordinates $\tau$ and $\vec x$ are dimensionless with this normalization. Next use 
\be
\hat a = \frac{\hat a_n}{\hat \tau_n - \tau} \, , ~~ {\rm for ~accelerated ~expansion} \, ; ~~~~~
\hat a = {\hat a_{n+1}}({\hat \tau_{n+1} + \tau}) \, , ~~ {\rm for ~radiation} \, ;
\label{scalef}
\ee
to describe the scale factors for the $n^{th}$ stage of acceleration and $(n+1)^{st}$ stage of radiation. 
These metrics need to be stitched together at time instants
\be
\tau_n = \tau_0 + \sum_{j=1}^{n} \Delta \tau_{j} \, ,
\label{tauk}
\ee
separating $n^{th}$ and $(n+1)^{st}$ stage, where $\Delta \tau_{n}$ are conformal time durations of individual epochs
comprising our atlas. 

We will take the convention
that even-numbered stages describe accelerated expansion and odd ones radiation eras. This is since we expect that while the
initial evolution prior to the rollercoaster dynamics is irrelevant, it should end with a short burst of accelerated expansion that sets up the 
BD state \cite{Kaloper:2018zgi} as the initial quantum state of rollercoaster cosmology. 
We will count this pre-rollercoaster epoch as the zeroth stage of the rollercoaster evolution. 
  
With these conventions, by substituting equations (\ref{scalef}) and (\ref{tauk}) into (\ref{bcs}) we obtain 
the recursion formulas relating the scale factors and the clock constants in adjacent stages, 
\ba
\hat \tau_{2n+1} &=& \hat \tau_{2n} - 2 \tau_{2n} \, , ~~~~~~~~~~~~~~~ \,  \hat a_{2n+1} = \frac{\hat a_{2n}}{[\hat \tau_{2n} -  \tau_{2n}]^{2}} \, ,
\nonumber \\
\hat \tau_{2n+2} &=& \hat \tau_{2n+1} + 2 \tau_{2n+1} \, , ~~~~~~~~~~ \hat a_{2n+2} = {\hat a_{2n+1}}[\hat \tau_{2n+1} + \tau_{2n+1}]^{2} \, .
\label{rec1}
 \ea 
We choose the first integration constant to be $\hat \tau_0 =1+\tau_0$, which fixes the gauge such that we
do not count the preinflation efolds towards the total tally of accelerated expansion. We also fix the length-scale gauge such that $\hat a_0 = 1$. 
In what follows we will use pre-rollercoaster and preinflation interchangeably, designating
the same thing: the additional epoch of accelerated expansion needed to prepare the initial state to be the BD vacuum. 

We can then
readily solve these recursion relations. First, Eqs. (\ref{rec1}) give $\hat \tau_1 = 1 - \tau_0$, $\hat a_1 = 1$. 
Next, a simple trick 
is to rewrite the recursion relations mutually connecting nearest accelerating stages with each other -- all with even $2n$ -- and nearest 
radiation stages -- counted by odd $2n+1$.
Using (\ref{tauk}) this yields 
\ba
&&\hat \tau_{2(n+1)} =\hat \tau_{2n} + 2 \Delta \tau_{2n+1}  \, , ~~~~~~~~~~~~  \, \hat a_{2(n+1)} = 
{\hat a_{2n}} \Bigl(\frac{\hat \tau_{2n+1} + \tau_{2n+1}}{\hat \tau_{2n} - \tau_{2n} } \Bigr)^2 \, , ~~~ \\
&&\hat \tau_{2(n+1)+1} = \hat \tau_{2n+1} - 2\Delta \tau_{2(n+1)} \, , \,\, \,~~~~\hat a_{2(n+1)+1} = 
{\hat a_{2n+1}}\Bigl(\frac{\hat \tau_{2n+1} + \tau_{2n+1}}{\hat \tau_{2n+2} - \tau_{2n+2} } \Bigr)^2 \, , ~~~~~~ \nonumber 
\label{rec2}
\ea
and so, after repeated application of these formulas we find the expressions for the time shift constants resetting the local clocks and the scale factor overall normalizations,
\ba
&&~~~~~~~~~~~ \hat \tau_{2n} = 1+\tau_0 + 2\sum_{i=0}^{n-1} \Delta \tau_{2i+1} \, , \nonumber \\
&&~~~~~~~~~~~ \hat \tau_{2n+1} = 1 -\tau_0 - 2 \sum_{i=1}^{n} \Delta \tau_{2i} \, , \nonumber \\
&&\hat a_{2n} = \prod_{j=1}^{n} \Bigl(\frac{1+\sum_{i=0}^{j-1} \Delta\tau_{2i+1} -\sum_{i=1}^{j-1} \Delta\tau_{2i}}{1+\sum_{i=0}^{j-2} \Delta\tau_{2i+1} -\sum_{i=1}^{j-1} \Delta\tau_{2i} } \Bigr)^2 \, , ~~~ \nonumber \\
&&\hat a_{2n+1} = \prod_{j=1}^{n} \Bigl(\frac{1+\sum_{i=0}^{j-1} \Delta\tau_{2i+1} -\sum_{i=1}^{j-1} \Delta\tau_{2i}}{1+\sum_{i=0}^{j-1} \Delta\tau_{2i+1} -\sum_{i=1}^{j} \Delta\tau_{2i} } \Bigr)^2 \, .
~~~~ \ea
Thus the metric quilt describing the background is
\be
ds^2 = 
\begin{cases}
\ldots \\
\frac{\hat a^2_{2j}}{[1+ \tau_0 +2 \sum_{i=0}^{j-1} \Delta \tau_{2i+1} - \tau]^2 } \frac{[-d\tau^2 + d\vec x^2]}{H^2} \, , \\
\hfill   \tau_0 +\sum_{i=1}^{2j-1} \Delta \tau_i  \le \tau \le \tau_0 +\sum_{i=1}^{2j} \Delta \tau_i   \, ;\\
\ldots \\
\hat a^2_{2n+1} \Bigl[1- \tau_0- 2\sum_{i=1}^{n} \Delta \tau_{2i}  + \tau\Bigr]^2 \frac{[-d\tau^2 + d\vec x^2]}{H^2} \, ,  \\
 \hfill ~~~~~~~~~~~~~~~~~~~~~~~~~~~~~~~~~~~~~~~~~~~~~~\tau_0 +\sum_{i=1}^{2n} \Delta \tau_i  \le \tau \le \tau_0 +\sum_{i=1}^{2n+1} \Delta \tau_i  \, ; \\
\ldots
\end{cases}
\label{metricquilt}
\ee
with the normalizing factors $\hat a_{n}$ given above. 
This is the most general solution describing the sequence of stages of accelerated expansion interrupted by decelerating radiative epochs,
all of arbitrary duration controlled by $\Delta \tau_i$.  

It is straightforward to check that during $ (2n)^{th}$ stage of accelerated expansion the scale factor increases by
\be
\frac{\hat a(\tau_{2n})}{\hat a(\tau_{2n-1})} = 
\frac{1+\sum_{i=0}^{n-1} \Delta\tau_{2i+1} -\sum_{i=1}^{n-1} \Delta\tau_{2i}}{1+\sum_{i=0}^{n-1} \Delta\tau_{2i+1} -\sum_{i=1}^{n-1} \Delta\tau_{2i} - \Delta \tau_{2n}} = e^{N_{2n}(I)} \, ,
\label{efoldsI}
\ee
whereas during $(2n+1)^{st}$ stage of radiation, the scale factor increases by 
\be
\frac{\hat a(\tau_{2n+1})}{\hat a(\tau_{2n})} =\frac{1+\sum_{i=0}^{n-1} \Delta\tau_{2i+1} 
-\sum_{i=1}^{n} \Delta\tau_{2i} + \Delta\tau_{2n+1}}{1+\sum_{i=0}^{n-1} \Delta\tau_{2i+1} -\sum_{i=1}^{n} \Delta\tau_{2i}}  = e^{N_{2n+1}(R)} \, ,
\label{efoldsR}
\ee
where we have used the logs of the scale factor variations $N_n$ as a measure of expansion for each stage. We can readily invert these equations
to get 
\ba
\Delta \tau_{2n} &=& \Bigl(1-  e^{-N_{2n}(I)}\Bigr) \Bigl(1+\sum_{i=0}^{n-1} \Delta\tau_{2i+1} -\sum_{i=1}^{n-1} \Delta\tau_{2i} \Bigr)  \, , \nonumber \\
\Delta \tau_{2n+1} &=& \Bigl(e^{N_{2n+1}(R)} - 1 \Bigr) \Bigl(1+\sum_{i=0}^{n-1} \Delta\tau_{2i+1} -\sum_{i=1}^{n} \Delta\tau_{2i} \Bigr)  \, .
\label{tauiter}
\ea
Once we pick an underlying theory where we calculate the expansion rates and duration of each epoch from first principles,
and thus determine $N_k$, we can easily compute the coordinate representation in conformal time by evaluating the conformal
time intervals $\Delta \tau_k$ recursively, following (\ref{tauiter}). In general, given the sequence of $N_{2n}(I), N_{2n+1}(R)$, we recognize 
(\ref{tauiter}) as a system of linear equations for $\Delta \tau_{2n}, \Delta \tau_{2n+1}$,
\ba
\Delta \tau_{2n} - \Bigl(1-  e^{-N_{2n}(I)}\Bigr) \Bigl(\sum_{i=0}^{n-1} \Delta\tau_{2i+1} -\sum_{i=1}^{n-1} \Delta\tau_{2i} \Bigr)  &=& \Bigl(1-  e^{-N_{2n}(I)}\Bigr) \, , \nonumber \\
\Delta \tau_{2n+1} - \Bigl(e^{N_{2n+1}(R)} - 1 \Bigr) \Bigl(\sum_{i=0}^{n-1} \Delta\tau_{2i+1} -\sum_{i=1}^{n} \Delta\tau_{2i} \Bigr)  &=& \Bigl(e^{N_{2n+1}(R)} - 1 \Bigr)  \, ,
\label{tauitermat}
\ea
with a lower triangular matrix, which is straightforward to invert to solve for  $\Delta \tau_{2n}, \Delta \tau_{2n+1}$ for any sequence of
$N_{2n}(I), N_{2n+1}(R)$. While seeming complex, all this -- as we noted above -- is just the construction of the single conformal time coordinate system covering the 
patches of the rollercoaster cosmology. 

For our purpose, it will suffice to consider the special case where all the epochs of accelerated and decelerated expansion yield the same
total expansion of the scale factor, respectively, so that $N_n(R) = N(R)$, $N_n(I) = N(I)$. Clearly more general cases are interesting, 
but in the interest of simplicity we will defer their analysis for later. In this case 
we can turn 
(\ref{tauiter}) into a recursion relation which is straightforward to solve:
\ba
\Delta \tau_{2n+1} &=& \Bigl(\frac{e^{N(R)}-1}{e^{N(I)}-1}\Bigr)  \Delta\tau_{2n} \, ,  \nonumber \\
\Delta \tau_{2n} &=& \Bigl(\frac{1-e^{-N(I)}}{1-e^{-N(R)}}\Bigr)  \Delta\tau_{2n-1} \, . \label{tauiternew}
\ea
We will employ these equations in the explicit examples below; they show that the sequences of conformal time durations 
of the epochs of accelerated and decelerated expansions of fixed scale factor growths represent geometric series in the conformal time coordinate. 

A particularly simple limit arises when all the epochs, both of accelerated expansion and of radiative interruptions are of the same conformal duration. 
When $\Delta \tau_{2n} = \Delta \tau_{2n+1} = \Delta \tau$, the metric quilt 
(\ref{metricquilt}) simplifies to 
\be
ds^2 = 
\begin{cases}
\ldots \\
\frac{(1+\Delta\tau)^{4j}}{[1+ \tau_0 +2j \Delta \tau - \tau]^2 } \frac{[-d\tau^2 + d\vec x^2]}{H^2} \, ,  
\hfill   \tau_0 +(2j-1) \Delta \tau \le \tau \le \tau_0 + 2j \Delta \tau \, ;\\
\ldots \\
(1+\Delta\tau)^{4n} \Bigl[1- \tau_0- 2n\Delta \tau + \tau\Bigr]^2 \frac{[-d\tau^2 + d\vec x^2]}{H^2} \, ,  ~~~  \hfill \tau_0 + 2n \Delta \tau 
 \le \tau \le \tau_0 +(2n+1) \Delta \tau \, . \\
\ldots 
\end{cases}
\label{metricquiltsim}
\ee
In this case the effective potential controlling the evolution of perturbations is especially simple, comprising of the equal number of identical
two-stage segments each of the same duration $2\Delta \tau$: one era of accelerated expansion and one of radiation domination,
each lasting $\Delta \tau$. As we will comment below, this case is of more limited phenomenological application, since the total number of
stages must not be too large in order to match to the size of the observed universe. Nevertheless, the inherent simplicity of this limit
will make the main claims very transparent. 

To begin to connect the rollercoaster idea to phenomenology, note that while the Hubble parameter during a stage of accelerated expansion remains approximately constant, it decreases during
the subsequent radiation change. We can easily find that the rate of change of the Hubble parameter between two nearest stages 
of cosmic acceleration is controlled by the rate of change of the scale factor during the radiation epoch that separates
them. Indeed, using the comoving Hubble parameter,
\be
H_{n+2} = H_n \Bigl(\frac{\hat a(\tau_n)}{\hat a(\tau_{n+1})} \Bigr)^2 \, .
\ee
This is precisely the purpose of the normalizing factor $\hat a_k^2$ in (\ref{metricquilt}): it accumulates the 
dilutions of the Hubble parameter during a stage of cosmic acceleration due to the preceding stages. 

How long should the individual stages of cosmic acceleration and deceleration be? We have already shown that a minimum of
$N \la 50-60$ efolds of accelerated expansion is required to solve the horizon problem. It may take a few more to also solve the flatness
problem, since the degeneracy of these two problems is now broken by the intervening early stages of radiation, during which 
$\Omega_k$ grows\footnote{However, the flatness problem may be at least partially alleviated during the pre-rollercoaster evolution.}. 
The main issue is however to ensure that the theory fits the observations of CMB and LSS. In other words,
the spectrum of fluctuations generated by rollercoaster dynamics must be roughly scale invariant over the range of scales on the sky where
it is observed to fit the Harrison-Zeldovich profile. The CMB data cover the range of scales from about ${\cal O}(10) \, {\rm GPc}$ to about 
${\cal O}(10) \, {\rm MPc}$, and the addition of LSS data widens the range to roughly ${\cal O}(100) {\rm \, kPc}$ or so. This means, to ensure that
the approximate scale invariance persists over this range of scales, and we imagine that the initial perturbations are imprints of quantum
fluctuations like in standard long inflation, we need to ensure that the Hubble parameter remains approximately constant over a scale factor variation of about 
\be
\frac{\hat a(\tau_n)}{\hat a(\tau_{n-1})}|_{CMB}  \simeq \frac{{\cal O}(10) \, {\rm GPc}}{{\cal O}(10) {\rm \, MPc}} \simeq 10^3 \, , ~~~~~~~~~ 
\frac{\hat a(\tau_n)}{\hat a(\tau_{n-1})}|_{CMB+LSS} \simeq \frac{{\cal O}(10) \, {\rm GPc}}{{\cal O}(100) {\rm \, kPc}} \simeq 10^5 \, , 
\ee
or therefore, $N_{CMB} \sim \ln 10^3 \sim 7$ to $N_{CMB+LSS} \sim \ln 10^5 \sim 11$ efolds. 

Moreover, after being imprinted onto the background, the fluctuations must remain
safely frozen at the largest scales in order to retain scale invariance until horizon reentry at very late times. 
In wave-mechanical language, the superhorizon perturbations behave like quantum-mechanical waves in
classically forbidden regions, with imaginary momenta. This means the barriers should be wide enough. In our case, we take
therefore the total accelerated expansion to come from several epochs with about equal, significant number of efolds each -- for example,
involving $4-6$ epochs of ${\cal O}(15)$ efolds apiece\footnote{In fact, if the intervening epochs of radiation are long we need additional 
efolds of accelerated expansion to keep the curvature small enough today, as we have seen thanks to Eq. (\ref{kinfl}). 
In the rest of the paper we will mostly ignore this correction, since we will imagine the radiation epochs to be shorter.}, at most. A more detailed analysis along the lines of \cite{dodhui,liddleleach,clifton} would be 
required to match to the data once a precise model is provided. We stress however that if the era of acceleration during which CMB anisotropies are generated is long enough -- longer than the minimal estimates above by the amount of expansion to at least compensate for the expansion during the subsequent radiation era -- the modes probed by the CMB will not reenter the horizon
immediately but will stay frozen until very late. We will see this explicitly in numerical examples in what follows. We note here that
recently similar mode behavior for gravity waves generated in phase transition was studied in \cite{johnco}. The power of those 
modes after horizon reentry behaves qualitatively the same as for modes which evolve through the rollercoaster. 

As we noted we separate acceleration eras by epochs of radiation. Overall, these epochs should not be too long in order
to make sure that the Hubble parameter at the end of the last stage of accelerated expansion remains higher than -- at the very least -- 
the Hubble parameter at nucleosynthesis. To play it safe, and minimize the modifications to the sector of cosmology controlled
by the Standard Model physics, we will in fact require that $H_{final}$ is greater than the Hubble factor at the QCD scale,
$H_{QCD} \simeq \Lambda_{QCD}^2/M_{Pl} \simeq 10^{-9} \, {\rm eV}$. If we take the initial Hubble parameter to be 
$H_{initial} \sim 10^{14} \, {\rm GeV}$, as close as possible to the bound imposed by the tensor-scalar ratio, and imagine that there 
were $4$ stages of accelerated expansion along with the initial stage of accelerated expansion that prepared the BD vacuum,
this means we have $4$ stages of early radiation epochs, during which the accumulated variation of the scale factor 
must not exceed $\sqrt{H_{initial}/H_{QCD}} \simeq 10^{16}$. 

These are the rough `boundary conditions' required to safely match the rollercoaster evolution 
to the data. It is clear from the construction that in principle
there could be many possible choices for the sequence of $N_{2k}(I), N_{2k+1}(R)$ that may lead to a realistic cosmology. Once one picks a model, more refined bounds along the lines of \cite{dodhui,liddleleach,clifton} should be employed. For simplicity
in what follows we will focus only on several special choices, where all epochs of accelerated expansion and all epochs of radiation
last the same, respectively. We will also consider, as an especially simple prototype, the case when we have two epochs of accelerated expansion 
each lasting $\sim 25$ efolds, which are preceded by one epoch of radiation, separated by another, and terminated with the third epoch
of radiation, all of which also last $\sim 25$ efolds. Clearly the first epoch will not matter, because it was succeeded by $50$ efolds of acceleration.
However, we will see that the perturbations generated during the first era of accelerated expansion will not be significantly
disturbed by the radiative interruption right after it. In order to show this, we now turn to setting the stage for the analysis of perturbations
on these backgrounds.

\section{Perturbations}

The standard analysis of perturbations generated during inflation involves computing the `backreaction' of quantized matter fields of a definite 
spin onto the background geometry using gauge-invariant perturbation theory around a quasi-de Sitter geometry. 
The quantization of the theory in the BD vacuum provides the normalization for the perturbation amplitude at the horizon crossing, where the local
excitations, behaving as harmonic oscillators at short wavelengths, transmute to the frozen curvature modes at superhorizon wavelengths,
because accelerated expansion takes the mode phases out of causal contact (see, e.g. the classic review \cite{Mukhanov:1990me}). The precise details are important to extract just how a quantum
fluctuation in the BD state affects the curvature. Yet, in all known cases of long inflation, the universal formulas for the perturbations emerge
regardless of the actual details. Since in this work we are principally pursuing the kinematic formulation of the rollercoaster cosmology 
including its ability to generate the correct spectrum of perturbations, with as little detail of the underlying microscopic theory as possible,
it is very useful to understand the perturbations in inflation from the universal, bird-eye's point of view. To this end, we return to the original,
augural work \cite{Mukhanov:1981xt}, which ushered the mechanism of quantum generation of perturbations into cosmology, and review it from
the vantage point of symmetries. We will find that all inflationary models can be understood as mere field redefinitions of the analysis
of \cite{Mukhanov:1981xt}, which will provide us with the framework to understand how perturbations form in rollercoaster cosmology. 

The authors of \cite{Mukhanov:1981xt} formulated the theory of the origin of perturbations using the model of 
\cite{Starobinsky:1980te}, which was based on a simple theory of semi-classical gravity. In addition
to the standard Einstein term, it also included the quadratic curvature term,
\be
S_{Starobinsky} = \int d^4x \sqrt{g} \Bigl(\frac{M_{Pl}^2}{2} R + c R^2 \Bigr) \, ,
\label{star}
\ee
where the dimensionless coefficient $c$ is a theory-dependent, in principle calculable, number. With $c \ga 0$ the theory is known
to have no instabilities \cite{Stelle:1976gc} and thus it can be viewed as a limit of a consistent QFT coupled to gravity. A key insight of
\cite{Mukhanov:1981xt} was to realize that although in the conventional GR there is no dynamical, propagating scalar field
in the metric sector, in the theory (\ref{star}) the extra scalar appears due to the quadratic curvature term. Realizing that
this scalar is buried in the metric itself, by inspecting equations of motion which yield $(\nabla^2 + m^2) R = 0$
with $m^2 \sim M_{Pl}^2/c$, the authors of \cite{Mukhanov:1981xt} proceeded to quantize it, and directly calculate its expectation value
in the long wavelength limit. Lo and behold, the scale invariant spectrum of perturbations popped out. Similar calculations can
be applied to the tensor curvature perturbations, to give the tensor spectrum of perturbations. The vector spectrum turns out to be
zero, by diffeomorphism and local Poincare symmetry. 

With some wisdom after the fact, we can readily understand why this apparent miracle occurred. First of all,  
in the regime of high curvatures, $R \ga M_{Pl}^2/c$, the terms with the highest number of derivatives dominate. So in the region
where the corrections smooth out the expected singularity \cite{Starobinsky:1980te}, the theory 
is dominated by the quadratic curvature term, 
\be
S_{Starobinsky} \rightarrow \int d^4x \sqrt{g} \, c \, R^2 \, .
\label{starsq}
\ee
At first sight, this theory seems to be very different from standard GR. It appears to have quartic terms in the gravitational equations
of motion, and it also has manifest scale invariance, where metric rescalings $g_{\mu\nu} \rightarrow \ell^2 g_{\mu\nu}$ do not
change the action. This symmetry gives a clue about the real nature of the theory, however. On any background with nonvanishing curvature,
the scaling symmetry is spontaneously broken. Then, by Goldstone theorem, the theory (\ref{starsq}) 
must have a massless Goldstone boson, which by local Poincare invariance must be a scalar. Finally, since the action (\ref{starsq}) 
is scale invariant, the background curvature $R = 12 H^2$ is the only scale. Since the theory has tensor 
fluctuations -- being a covariant theory of the metric -- their coupling, which defines the effective Planck scale on the background of curvature $H$, must be $M_{Pl}^2({\rm eff}) \sim c H^2$.
The $c$ dependence follows from the decoupling limit, $M_{Pl}({\rm eff}), c \rightarrow \infty$, which freezes the tensor fluctuations.
This also shows that in the limit $H \rightarrow 0$ where the scaling symmetry remains unbroken, the gravitational theory 
(\ref{starsq}) is in strong coupling, $1/M_{Pl}({\rm eff}) \rightarrow \infty$. Hence the action (\ref{starsq}) really must represent a 
one-parameter family of theories of gravity with one given scale -- the background curvature scale -- and a massless scalar field,
\be
\int d^4x \sqrt{g} \, c \, R^2 \equiv \int d^4x \sqrt{\tilde g}  \Bigl( \frac{M^2_{Pl}({\rm eff})}{2} \tilde 
R - \frac12 (\tilde \nabla \phi)^2 - \Lambda({\rm eff}) \Bigr) \, , \label{unigr}
\ee
where
\be
M_{Pl}({\rm eff})^2 = 48  c  H^2 \, , ~~~~~~~~~ \Lambda({\rm eff}) = 144  c  H^4 \, , 
\label{dimc}
\ee
and $H$ plays the role of an arbitrary scaling symmetry breaking spurion. 

The addition of the Einstein's term to the quadratic theory
modifies this story only slightly, since it adds explicit scaling symmetry breaking via the nonvanishing 
Planck scale in the flat space limit. This ensures that the theory remains weakly coupled at all curvatures, gives an explicit mass 
term to the scalar, making it a pseudo-Goldstone mode instead, fixes the particular value of $H$ in the scale-invariant
limit where the quadratic term dominates, and also renders the background curvature explicitly dependent on the Starobinsky scalaron,
a.k.a. the broken scaling symmetry pseudo-dilaton. A full analysis of the details of these quadratic curvature theories can be found in 
\cite{Stelle:1976gc,Alvarez-Gaume:2015rwa,Ferrara:2015ela,Csaki:2014bua}. Here we will only consider the transformation that yields
(\ref{unigr}), which corresponds to taking the theory (\ref{starsq}) to the axial gauge. This will suffice for our purposes,
as we shall show shortly.

To prove Eq. (\ref{unigr}), we first rewrite (\ref{starsq}) as a scalar-tensor theory using two Lagrange multipliers,
\be
S^{2} = \int d^4x \sqrt{g} \Bigl( c \Phi R + \lambda (\Phi - R) \Bigr) \, .
\label{scte}
\ee
This action is exactly the same as (\ref{starsq}) which can be seen by integrating $\lambda$. Next define
$\lambda = c \Phi - c \chi$, rewrite (\ref{scte}) as $c \chi R + c (\Phi - \chi) \Phi$ and integrate out
$\Phi$ using $\Phi = \chi/2$. This yields $\int d^4x \sqrt{g} c [\chi R - \chi^2/4]$. Finally, carry out the 
conformal transformation and field redefinition
\be
g_{\mu\nu} = \frac{\chi_0}{\chi} \tilde g_{\mu\nu} \, , ~~~~~~~~~~~~ \chi = \chi_0 \exp(\frac{\phi}{\sqrt{3c\chi_0}}) \, , 
\label{conftra}
\ee
where $\chi_0$ is a completely arbitrary dimensional scale, of dimension $({\rm mass})^2$. This yields
$\int d^4x \sqrt{\tilde g}[ c\chi_0 \tilde R - (\tilde \nabla \phi)^2/2 - c\chi_0^2/4]$, and thus 
\be
M_{Pl}({\rm eff})^2 = 2 c \chi_0 \, , ~~~~~~~~~~~ \Lambda = c \frac{\chi_0^2}{4} \, . ~~~~~~~
\label{sclas}
\ee
Evaluating the curvature on the background given by scales (\ref{sclas}) yields $H^2 = M_{Pl}^2({\rm eff})/48c$, and therefore the
final result is precisely (\ref{unigr}) with (\ref{dimc}). 

This result is very useful since it shows that any slow roll theory of inflation, which to the leading -- zeroth -- order in slow roll parameters is
given precisely by Eq. (\ref{unigr}), is isomorphic to the quadratic curvature action (\ref{starsq})! This immediately shows why the 
perturbations are scale invariant: this follows from the global scale invariance of the action (\ref{starsq}), spontaneously broken by the 
nonvanishing background curvature, which leads to the imprint of the Goldstone fluctuations onto the background geometry.
The mechanism which realizes this is just a limiting form of the original setup by Mukhanov and Chibisov, and may be viewed as the simplest effective theory of  inflationary fluctuations~\cite{Cheung:2007st,Weinberg:2008hq}.
The challenge in constructing inflationary models, then, is finding mechanisms that lead to an 
emergent, approximate scaling symmetry which is maintained through a sufficiently large swath of the spacetime. This is a challenge in QFT 
and perhaps even more in quantum gravity, which has led to many different approaches to date to implement it. However essentially
all of the approaches, particularly those that attempt to generate a long smooth inflation, are equivalent to the leading order  to the
scale invariant limit of Starobinsky theory, endowed with the Mukhanov-Chibisov method for generating perturbations. 

Conversely, this analysis immediately tells us that whenever we encounter a theory which supports a regime where 
the action is very nearly scale invariant, a stage of inflation will result which will generate a scale-invariant spectrum of perturbations. Clearly
the scale invariance cannot be exact if the now-standard lore that quantum gravity prohibits exact global symmetries is right \cite{bs}. Thus
the field perturbations will imprint onto the metric perturbations, and to find the spectrum all we need to do is to map (\ref{unigr}), (\ref{dimc}) 
back on (\ref{starsq}) and import the formulation of Mukhanov and Chibisov for each dynamical regime where the action is scale invariant. That's it. 

This is the approach we take here. We imagine that the stages of accelerated expansion in the rollercoaster cosmology are approximately
de Sitter with a scalar in very slow roll, and that there is one order parameter controlling scaling symmetry, which plays the role of the quasi-dilaton
for each stage. The effective dynamics of the model in such a stage is then given by (\ref{unigr}), which in turn is equivalent to (\ref{starsq}). 
This means, we interpret both the background evolution and the fluctuating field as the purely gravitational sector, and apply directly the
approach motivated by \cite{Mukhanov:1981xt}. Since on shell $R = \Phi = \chi/2 = (\chi_0/2)  \exp(\frac{\phi}{\sqrt{3c\chi_0}})$ we immediately find that
the quasi-dilaton fluctuations are related to the fluctuations of the scalar curvature in (\ref{starsq}) by $\frac{\delta R}{R} = \sqrt{\frac23} \frac{\delta \phi}{M_{Pl}}$, or, substituting $R = 12 H^2$, 
\be
\delta \phi = \sqrt{\frac{c}{2}} \frac{\delta R}{H} \, , 
\label{deltar}
\ee
which is a special limit of the result of  \cite{Mukhanov:1981xt}, restricted to purely quadratic gravity action. To evaluate it, we used the formula for
the effective Planck mass in (\ref{dimc}). We can view this equation as a {\it de facto} definition of the ultralight quasi-dilaton whose
quantum fluctuations produce curvature perturbations in each stage of accelerated expansion during rollercoaster evolution. The
important point here is that during each stage of cosmic acceleration there is one scalar field, a `goldilocks' which controls the rate of decay of 
transient vacuum energy. This field feeds into the curvature and transfers to it its fluctuations, by equivalence principle. There could be other
scalar fields; but if they are heavy they decouple; if they are much lighter, during the epoch of cosmic acceleration they are slowed down
much more efficiently than the `goldilocks' field, and even though their fluctuations are the same, $\langle \delta \phi^2 \rangle \sim H^2$,
their contribution to the curvature is suppressed. They could lead to problems if they are ultralight, in which case their very late decay could 
yield isocurvature perturbations. Further, lighter fields will not disturb the CMB as long as the stage during which CMB anisotropies are
formed is long enough, as we discussed in the previous section. We will therefore ignore isocurvature perturbations in the rest of this work. 

The derivation of the
tensor fluctuations is even more straightforward, since the theory contains fundamental tensor modes -- i.e. metric fluctuations. Each of the two behaves precisely as a massless scalar field. 

Thus all the modes to the leading order -- where we approximate evolution as being scale invariant -- propagate as exactly massless scalars, with the action
\be
S_{scalar} = - \int d^4x \sqrt{\tilde g} \frac12 (\tilde \nabla \phi)^2 \, .
\label{scalact}
\ee
Working with spatially flat FRW coordinates, to eventually Fourier-transform spatial coordinates thanks to manifest translational
symmetries in spatial directions, we find that the three helicities, after redefinition $\varphi = a \phi$, are all described by the action
\be
    S_{scalar} = \int \rmd \tau \rmd^3 x \[ (\vphi')^2 - (\vec \nabla \varphi)^2  + \frac{z''}{z} \varphi^2 \] \, ,
\label{scalaracts}
\ee
where we now employ the conformal time introduced in the previous section, for convenience. In the case of the helicity-0,
we really use the gauge-invariant Mukhanov-Sasaki variable $\vphi = a \phi - \frac{a \phi'}{\calH} \Psi$, where
$\Psi$ is the metric perturbation representing Newtonian potential in the longitudinal gauge \cite{Mukhanov:1981xt,Sasaki:1986hm}. 
Here $z = \frac{a \phi'}{\calH} = \frac{a \dot{\phi}}{H}$, and in the scale-invariant limit -- i.e. to the leading order in slow roll parameters -- $\frac{z''}{z} = \frac{a''}{a}$.

Quantizing (\ref{scalaracts}) is straightforward. Fourier-transforming and defining canonical variables 
\begin{equation}
    \vphi_{\vk}(\tau) = \int \frac{\rmd^3 x}{(2\pi)^{3/2}} e^{- i \vk \cdot \vx} \vphi(\tau, \vx) \, , \quad \quad
    \pi_{\vk}(\tau) = \int \frac{\rmd^3 x}{(2\pi)^{3/2}} e^{- i \vk \cdot \vx} \pi(\tau, \vx) \, ,
    \label{momcoords}
\end{equation}
which satisfy the reality condition $\vphi_{\vk}^\dag = \vphi_{-\vk}$, $\pi_{\vk}^\dag = \pi_{-\vk}$, and the canonical relation
$\pi_k = \varphi_k'$, as is obvious from (\ref{scalaracts}), we can write the 
mode expansion of the field $\vphi(\tau, \vx)$ (using $k=|\vec k|$):
\begin{equation}
    \vphi(\tau, \vx) = \int \frac{\rmd^3 k}{(2 \pi)^{3/2}} \[ b(\vk) u_k(\tau) e^{i \vk \cdot \vx} +
    b^\dag(\vk) u_k^\ast(\tau) e^{- i \vk \cdot \vx} \] \, .
    \label{phifour}
\end{equation}
The annihilation and creation operators $b(\vk)$, $b^\dag(\vk)$ satisfy the usual commutation relations
\begin{equation}
    \comm{b(\vk)}{b^\dag(\vq)} = \delta^{(3)}(\vk-\vq) \, , \quad 
    \comm{b(\vk)}{b(\vq)} = \comm{b^\dag(\vk)}{b^\dag(\vq)} = 0 \, ,
    \label{eq:comm}
\end{equation}
with the mode wavefunctions\footnote{From here on, we will  refer to these was ``wavefunctions" for 
reasons of simplicity, but also because in the linearized limit they are in fact precisely Fourier transforms
of the solutions of wave equations in FRW spacetime. We find this useful as it allows for direct
application of wave mechanics intuition in what follows. Formally, one could just use the term  "mode functions" instead.}
defined as the solutions of  
\begin{equation}
    u_k'' + \(k^2 - \frac{a''}{a} \) u_k = 0 \, .
    \label{eq:maineq}
\end{equation}
In each epoch of radiation domination, this yields a vanishing effective potential for the fluctuations, since $a'' = 0$. 
The individual mode functions are just harmonic oscillators. In the epochs of accelerated expansion, approximated by
de Sitter metric, $a''/a = 2/(\hat \tau_{2k} - \tau)^2$, which means that the potential for fluctuation is a Bessel function
barrier in time. In fact we can evaluate it exactly using the formula for the rollercoaster quilt (\ref{metricquilt}). We find
\be
\frac{a''}{a} = 
\begin{cases}
\ldots \\
\frac{2}{[1+ \tau_0 +2 \sum_{i=0}^{j-1} \Delta \tau_{2i+1} - \tau]^2 }  \, ,
\hfill  ~~~~~  \tau_0 +\sum_{i=1}^{2j-1} \Delta \tau_i  \le \tau \le \tau_0 +\sum_{i=1}^{2j} \Delta \tau_i   \, ;\\
\ldots \\
~~~~~~~~~~~~~ 0 \, ,  \, ~~~~~~~~~~~~~~~~~~~\tau_0 +\sum_{i=1}^{2n} \Delta \tau_i  \le \tau \le \tau_0 +\sum_{i=1}^{2n+1} \Delta \tau_i  \, ; \\
\ldots
\end{cases}
\label{pertquilt}
\ee
Similar potentials -- for bulk modes -- were encountered in the crystal braneworlds in \cite{Kaloper:1999et}. 

What remains is to solve the Schr\"odinger equation \eqref{eq:maineq} for each individual stage,
with the potential given in Eq. (\ref{pertquilt}) using the appropriate boundary conditions for the mode functions
and stitch the solutions for different epochs together relating them with the perturbed Israel junction conditions, following the
examples of \cite{Deruelle:1995kd,Kaloper:2003nv}. Finding the solutions for each individual stage is straightforward for
(\ref{pertquilt}); they are given by 
\be
u_k = 
\begin{cases}
\ldots \\
\frac{1}{\sqrt{2k}} \Bigl[ A^{(j)}_k \Bigl(1 + \frac{i}{k (\hat \tau_j - \tau)} \Bigr) e^{-ik\tau} + B^{(j)}_k \Bigl(1 - \frac{i}{k (\hat \tau_j - \tau)} \Bigr) e^{ik\tau} \Bigr] \, , 
~~~~{\rm in ~j^{th}~accelerated ~era}    \, ;\\
\ldots \\
~~~~~~~~~~~~~~~~~~~ \frac{A^{(n)}_k}{\sqrt{2k}} e^{-ik\tau} +  \frac{B^{(n)}_k}{\sqrt{2k}}  e^{ik\tau} \, ,   ~~~~~~~~~~~~~~~~~~~~~~~~~{\rm in ~n^{th}~radiation ~era}  \, . \\
\ldots
\end{cases}
\label{wavefctns}
\ee
Note that we have absorbed the clock synchronizing phases $e^{\pm i k \hat \tau_n}$ into the overall factors $A^{(n)}_k, B^{(n)}_k$. 
We have however kept the terms $\propto k \hat \tau_j$ -- as we must -- in the Hankel function prefactors in the top line of (\ref{wavefctns}). 
The potential $a''/a$ in (\ref{pertquilt}) is discontinuous, but the discontinuity is finite: because of this, preserving
unitarity at every transition from accelerated expansion to radiation and vice versa requires
\be
u_k(\tau_-) = u_k(\tau_+) \, , ~~~~~~~~~ u'_k(\tau_-) = u'_k(\tau_+) \, .
\label{bcsu}
\ee
Nonintegrable discontinuities can appear in the effective potential for perturbations $z''/z$, which includes $a''/a$ plus slow
roll corrections, which can lead to additional mode rotations on the interfaces, as shown in \cite{Kaloper:2003nv}.
One may be concerned about discontinuities such as these inducing large corrections to the evolution 
due to gravitational particle production, see e.g. \cite{ford,prok}, which technically amounts to vacuum mode function 
``squeezing".
However in real situations the transition in $a$ is smooth, and these effects generically yield energy densities of at most $\sim H^4$ per species.
Note that in models with many light species these effects may have significant consequences.
Since we assume that the number of light species here is not dramatically large, we will neglect these effects here. 

Finally, we need to specify the initial wavefunction, to be used in the first epoch of accelerated expansion in the
rollercoaster. This choice is fixed with the choice of the initial state. Having chosen the BD vacuum as the initial state at the
onset of the rollercoaster evolution corresponds to choosing the initial wavefunction with fixed spatial momentum $k$ 
to be the mode of only positive frequency coming out of the stage of preinflation which sets the BD state. 
This describes a mode propagating causally, moving to the future\footnote{This comes about as follows: the BD state is defined by a limiting process where the initial preparatory stage is maximally extended to the past, $\tau \rightarrow - \infty$. In FRW flat cover of de Sitter space, this
limit takes one to the past horizon of the spatially flat future patch. Any mode which is moving back to the past in the local free 
falling frame would experience huge blueshift and backreact on the geometry too much; thus the BD state must not contain any excitations 
flowing back in time in order to trust the QFT in it with arbitrary precision. One may relax this by ending the geometry sooner rather than later, but 
if after the cutoff a stage of acceleration occurs for ${\cal O}(10)$ efolds or more, the surviving distortions are irrelevant for late time physics \cite{Kaloper:2018zgi}. This is the
approach we follow here.}. Thus
\be
u_k(\tau)^{initial} =\frac{1}{\sqrt{2k}} \Bigl(1 + \frac{i}{k(1+\tau_0- \tau)} \Bigr) e^{-ik(\tau-\tau_0)} 
  \, .
\label{bcsin}
\ee
With this, we see that finding the wavefunction describing the mode evolution through the rollercoaster
is completely analogous to solving the Schr\"odinger equation for a propagation through a multibarrier system,
which behave as a block of material. This is almost like tunneling, with one important distinction: the initial wave
function (\ref{bcsin}) involves only the incident wave, and not the reflected wave from the first barrier. Thus we 
will refer to this dynamics as anti-tunneling. We will next solve the equation, and then use the net amplitude 
to determine the power spectrum of the propagated perturbations, which is controlled by the
$|{\rm amplitude}|^2$. 

\section{Wavefunctions and Their Power} 

We can now turn to solving for the wavefunctions $u_k$ defined in the previous section. We will approach this task in a modular
manner, devising the elementary building blocks for constructing the total wavefunction in steps that can be readily generalized
to any desired sequence of epochs. From inspecting the solution (\ref{metricquilt}), it is clear that the elementary steps are
\begin{itemize}
\item{1)} transition from the last pre-rollercoaster stage to the radiation stage afterwards;
\item{2)} transition from an arbitrary radiation stage to a stage of accelerated expansion;
\item{3)} and finally, transition from an arbitrary stage of accelerated expansion to a radiation stage.
\end{itemize}
Then any rollercoaster cosmology can be constructed as a sequence of transitions between various stages, which go as 
$$1) \, \rightarrow \, 2) \, \rightarrow \, 3) \, \rightarrow \, 2) \, \rightarrow \, \ldots \, \rightarrow \, 3) \, \rightarrow \, 2) \, \rightarrow \, 3)$$ 
where the durations of individual stages are inputs from a particular model, that ought to be constructed 
from first principles. Below we will first construct the rules for linking the wavefunctions for the three `modular' cases, and determine the general formula for constructing a wavefunction for any rollercoaster
cosmology by combining these. Finally we will develop the general expression for the power of a mode with a given 
wavefunction, and define the mode {\it form factor}, as a ratio of the exact solution to the expression for long inflation at fixed
spatial momentum $\vec k$ in order to make the comparison transparent.

\subsection{Wavefunction Refraction}

The matching of the wavefunctions from pre-rollercoaster to the first radiation stage is quite straightforward. Given the wavefunction (\ref{bcsin}) and the boundary conditions (\ref{bcsu}) at $\tau = \tau_0$, we need 
to calculate the coefficients $A_k, B_k$ representing the wavefunction in the first intervening radiation epoch, given by the second line in
Eq. (\ref{wavefctns}). The calculation is quite straightforward, yielding
\be
A_k = \Bigl(1 + \frac{i}{k} - \frac{1}{2k^2}\Bigr) e^{ik \tau_0}  \, , ~~~~~~~~~~~~ B_k = \frac{1}{2k^2} e^{-ik \tau_0} \, ,
\label{firstcoefs}
\ee
such that the wavefunction in the first radiation stage is given by 
\be
u_k^{(1)} =\frac{1}{\sqrt{2k}} \Bigl[ \Bigl(1 + \frac{i}{k} - \frac{1}{2k^2}\Bigr) e^{-ik(\tau-\tau_0)} 
+ \frac{1}{2k^2} e^{ik(\tau-\tau_0)} \Bigr] \,, ~~~~~~~ \tau_0 < \tau < \tau_0 + \Delta \tau_1 \, .
\label{radw1}
\ee
Note that the mode function has developed a negative frequency tail for any momentum $k$, due to the anti-tunneling 
boundary conditions imposed by the choice of the BD vacuum as the quantum state at the end of the pre-rollercoaster 
evolution, and that this happens to waves of arbitrary wavelength. This is not surprising: regardless of whether a wave
propagates above a barrier or through it, it scatters on it. 

However the specificity of the detail in (\ref{radw1}) obscures the elegance of the more general case, where 
we consider matching general wavefunctions in two adjacent epochs. First of all, we can see from (\ref{wavefctns}) that
in either accelerated or radiation stages, the mode functions are conjugates of each other. This obviously follows
from the fact that the Schr\"odinger equation (\ref{eq:maineq}) is Hermitean. Thus in any epoch we can write the 
most general wavefunction as a linear combination
\be
u_{k}^{(n)}(\tau) = A_k f^{(n)}_k(\tau) + B_k f_k^{(n)~*}(\tau) \, ,
\ee
where $f^{(n)}_k, f^{(n)~*}_k$ are either of the modes in (\ref{wavefctns}). Further, the modes are mutually orthogonal
in the standard Hilbert space metric defined by the Schr\"odinger equation (\ref{eq:maineq}), which is locally guaranteed by the fact that  
their Wronskian is nonzero,
\be
W = \det \begin{pmatrix}
f^{(n)}_k & f^{(n)~*}_k \\
f_k^{(n)}{}' & f_k^{(n)~*}{}' 
\end{pmatrix} = i \, .
\label{wronski}
\ee
The boundary conditions between $n^{th}$ and the $(n+1)^{st}$ stage (\ref{bcsu}) can then be written in the matrix form as (here the $f$'s are all evaluated at $\tau_n$)
\be
\begin{pmatrix}
f^{(n+1)}_k & f^{(n+1)~*}_k \\
f_k^{(n+1)}{}' & f_k^{(n+1)~*}{}' 
\end{pmatrix} 
\begin{pmatrix}
A^{(n+1)}_k  \\
B_k^{(n+1)} 
\end{pmatrix} = 
\begin{pmatrix}
f^{(n)}_k & f^{(n)~*}_k \\
f_k^{(n)}{}' & f_k^{(n)~*}{}' 
\end{pmatrix} 
\begin{pmatrix}
A^{(n)}_k  \\
B_k^{(n)} 
\end{pmatrix} \, , 
\label{matrixeqs}
\ee
which can be inverted to yield
\be
\begin{pmatrix}
A^{(n+1)}_k  \\
B_k^{(n+1)} 
\end{pmatrix} = - i 
\begin{pmatrix}
 f_k^{(n+1)~*}{}' & - f^{(n+1)~*}_k \\
- f_k^{(n+1)}{}' &  f^{(n+1)}_k 
\end{pmatrix} 
\begin{pmatrix}
f^{(n)}_k & f^{(n)~*}_k \\
f_k^{(n)}{}' & f_k^{(n)~*}{}' 
\end{pmatrix} 
\begin{pmatrix}
A^{(n)}_k  \\
B_k^{(n)} 
\end{pmatrix} \, , 
\label{matrixbcs}
\ee
where we used $W^{-1} = - i$ in the inversion of (\ref{matrixeqs}). We can readily check that (\ref{firstcoefs}), (\ref{radw1}) are a special case of this
matrix equation, applied to the initial state vector $A_k^{(0)} = 1, B_k^{(0)} = 0$. So if we define the unitary transition matrix $S$ 
\be
S^{(n+1), (n)} = - i 
\begin{pmatrix}
 f_k^{(n+1)~*}{}' & - f^{(n+1)~*}_k \\
- f_k^{(n+1)}{}' &  f^{(n+1)}_k 
\end{pmatrix} 
\begin{pmatrix}
f^{(n)}_k & f^{(n)~*}_k \\
f_k^{(n)}{}' & f_k^{(n)~*}{}' 
\end{pmatrix} 
\label{smatrix}
\ee
it is obvious that the state vector after $n$ transitions is
\be
\begin{pmatrix}
A^{(n+1)}_k  \\
B_k^{(n+1)} 
\end{pmatrix} = {\cal T} \Big\{ \prod_{j=0}^{n} S^{(j+1), (j)} \Bigr\}  
\begin{pmatrix}
1 \\
0 
\end{pmatrix} \, , 
\label{matrixbcsoln}
\ee
where ${\cal T}\Bigl\{\ldots\Bigr\}$ designates chronological ordering, placing matrices with larger $j$ later in the sequence of multiplications.
To work out the precise details all we need to do is count off the number of stages, pick their individual durations, and evaluate the wavefunctions
and their derivatives at end points of each stage. This defines the $S$-matrices in (\ref{smatrix}) and their chronological product in 
(\ref{matrixbcsoln}).

It is not practical to evaluate the matrix ${\cal T} \Big\{ \prod_{j=0}^{n} S^{(j+1), (j)} \Bigr\}$; however we can straightforwardly determine its building blocks,
i.e. $S$-matrices governing the transition from an accelerating to a radiation dominated epoch $S^{(2n+1), (2n)}$ and the converse, from
a radiation epoch to an epoch of accelerated expansion $S^{(2n), (2n-1)}$. The former occurs at the instant of time $\tau_{2n}$ defined in the
previous section, and is given by
\be
S^{(2n+1), (2n)} = \begin{pmatrix}
1 + \frac{i}{k(\hat \tau_{2n} - \tau_{2n})} -  \frac{1}{2 k^2(\hat \tau_{2n} - \tau_{2n})^2} 
~ & ~ \frac{1}{2k^2(\hat \tau_{2n} - \tau_{2n})^2} e^{2ik\tau_{2n}} \\
\frac{1}{2k^2(\hat \tau_{2n} - \tau_{2n})^2} e^{-2ik\tau_{2n}} ~&~ 1 - \frac{i}{k(\hat \tau_{2n} - \tau_{2n})} - \frac{1}{2k^2(\hat \tau_{2n} - \tau_{2n})^2} 
\end{pmatrix} \, ,
\label{smatone}
\ee
whereas the matrix $S^{(2n), (2n-1)}$ can be obtained from this expression by Hermitean conjugation and
replacement $\tau_{2n} \rightarrow \tau_{2n-1}$ (but without changing $\hat \tau_{2n}$!). The expression is therefore
\be
S^{(2n), (2n-1)} = \begin{pmatrix}
1 - \frac{i}{k(\hat \tau_{2n} - \tau_{2n-1})} -  \frac{1}{2k^2(\hat \tau_{2n} - \tau_{2n-1})^2} 
~ & ~ \frac{1}{2k^2(\hat \tau_{2n} - \tau_{2n-1})^2} e^{2ik\tau_{2n-1}} \\
\frac{1}{2k^2(\hat \tau_{2n} - \tau_{2n-1})^2} e^{-2ik\tau_{2n-1}} ~ &~ 1+ \frac{i}{k(\hat \tau_{2n} - \tau_{2n-1})} -  \frac{1}{2k^2(\hat \tau_{2n} - \tau_{2n-1})^2} 
\end{pmatrix} \, .
\label{smattwo}
\ee
Using a chronological product of these two basic blocks we can solve for the wavefunction in any rendition of rollercoaster cosmology. 

The important takeaway point is that the junction $S$-matrices, and in general the local evolution $S$-matrices,  are not diagonal. 
The off-diagonal elements, specifically, ensure that evolution
of the wave through the cosmic expansion generates a small negative frequency correction to an incident positive frequency wave no matter what.
This is of course the phenomenon of particle production in a background gravitational field: a state that is empty of particles at one time 
will contain them at another time since the time evolution rotates the field coordinates and momenta in the phase space. 
This is the core of the idea of generating metric perturbations from quantum fluctuations during inflation. It works the same in rollercoaster cosmology, since after all
the rollercoaster includes a sequence of many microinflations. 

\subsection{Wave Power} 

To calculate the power in the waves generated during inflation and their imprint on the geometry, we can use the Heisenberg operators of the theory,
where the time evolution is included into the operators $b, b^\dagger$ instead of in wavefunctions. Having solved for the wavefunctions already
determines the precise form of the wave operators. Indeed, the key point is to notice that the wavefunctions all reduce to the simple plane waves
in the limit $\tau \rightarrow - \infty$ in the preinflation stage, which sets up the BD state of the universe -- which in this limit reduces
to the local Minkowski space. This formalism extends to the rollercoaster framework, as we will now show. 

Using this and the known form of the wavefunctions, we can immediately find the Heisenberg operators
$b(\tau), b^{\dagger}(\tau)$ and the relation between these operators at two different times $\tau$ and $\tau'$. Since
\be
b_k(\tau) = \frac{1}{\sqrt{2}} \Bigl(\sqrt{k} \varphi_k(\tau) + \frac{i}{\sqrt{k}} \pi_k(\tau) \Bigr) \, , ~~~~~~~~~~
b^{\dagger}_k(\tau) = \frac{1}{\sqrt{2}} \Bigl(\sqrt{k} \varphi_{-k}(\tau) - \frac{i}{\sqrt{k}} \pi_{-k}(\tau) \Bigr) \, ,
\ee
where $\varphi_k, \pi_k$ are the Fourier transformed field and its conjugate momentum, respectively. Using the form of the solution for
$\varphi_k(\tau) = b(\vk) u_k(\tau) +  b^\dag(-\vk) u_k^\ast(\tau)$ which follows from (\ref{phifour}), and noting that 
$\pi_k = \varphi_k'$ gives 
\be
b_k(\tau) = {\cal F}_k(\tau) b_k + {\cal G}^*_k(\tau)b^\dagger_{-k}  \, , 
~~~~~~~~~~~ b_k^\dagger(\tau) = {\cal F}^*_k(\tau)  b^\dagger_k + {\cal G}_k(\tau)  b_{-k}\, .
\label{bsbog}
\ee
and the functions ${\cal F}, {\cal G}$ are 
\be
{\cal F}_k(\tau) = \frac{1}{\sqrt{2}} \Bigl(\sqrt{k} u_k(\tau) + \frac{i}{\sqrt{k}} u_k'(\tau)\Bigr) \, , ~~~~~~~~~ {\cal G}_k(\tau) = \frac{1}{\sqrt{2}} \Bigl(\sqrt{k} u_k(\tau) - \frac{i}{\sqrt{k}} u_k'(\tau) \Bigr)  \, .
\label{bsbogf}
\ee
It is straightforward to check that as $\tau \rightarrow -\infty$, ${\cal F}_k \rightarrow \exp(-i k \tau)$, ${\cal G}_k(\tau) \rightarrow 0$.
This means that the Schr\"odinger picture operators $b_k, b_k^\dagger$ can be interpreted as the initial conditions for the
Heisenberg operators $b_k(\tau), b_k^\dagger(\tau)$. Hence the equation (\ref{bsbog}), (\ref{bsbogf}) are in fact the solutions
for the Heisenberg equations of motion $iQ' = [Q,{\cal H}]$ for the second quantization operators. One can readily check that
${\cal F}_k {\cal F}_k^* - {\cal G}_k {\cal G}_k^* = 1$, and thus (\ref{bsbog}) is a Bogoliubov rotation of operators 
representing evolution from $\tau \rightarrow - \infty$ to $\tau$. Inverting and applying twice, we can write the evolution equation
propagating operators $b_k(\tau), b_k^\dagger(k)(\tau)$ from $\tau$ to $\bar \tau$; in matrix form,
\be
\begin{pmatrix}
b_k( \tau)  \\
b^\dagger_{-k}( \tau)
\end{pmatrix} = 
\begin{pmatrix}
{\cal M}_k(\tau,\bar \tau) &{\cal N}_k(\tau,\bar \tau)  \\
{\cal N}^*_k(\tau,\bar \tau) & {\cal M}^*_k(\tau,\bar \tau) 
\end{pmatrix} 
\begin{pmatrix}
b_k(\bar \tau)  \\
b^\dagger_{-k}(\bar \tau)
\end{pmatrix} \, , 
\label{matrixbog}
\ee
where
\be
{\cal M}_k(\tau,\bar \tau) = {\cal F}_k(\tau) {\cal F}_k^*(\bar \tau) - {\cal G}^*_k(\tau) {\cal G}_k(\bar \tau) \, , ~~~~~~~~~
{\cal N}_k(\tau,\bar \tau) = {\cal G}_k^*(\tau) {\cal F}_k(\bar \tau) - {\cal F}_k(\tau) {\cal G}_k^*(\bar \tau) \, .
\ee
Since the matrix in (\ref{matrixbog}) is a product of two Bogoliubov rotations, it must also be a Bogoliubov rotation. Hence specifically
\be
{\cal M}_k(\tau,\bar \tau) {\cal M}^*_k(\tau,\bar \tau) - {\cal N}_k(\tau,\bar \tau) {\cal N}^*_k(\tau,\bar \tau)  = 1 \, ,
\label{bogol}
\ee
with the other standard formulas constraining other matrix elements as well.

Note that all of these formulas, normally derived for smooth long inflation, remain applicable in the rollercoaster
cosmology case since unitarity is preserved at the junctions. Here this simply means that the evolution is adiabatic 
during individual accelerated expansion and radiation epochs, and obeys the unitary junction rules given by
$S$-matrices of Eqs. (\ref{smatrix}), (\ref{matrixbcsoln}), (\ref{smatone}), (\ref{smattwo}). We will discuss this important point 
in more detail shortly since it should be demonstrated clearly and explicitly. 

This sets up the required formalism for computing the power deposit of each wave in the BD state, which is formally defined by the
adiabatic condition
\be
b_k(\bar \tau) | 0(\bar \tau) \rangle = 0 \, .
\label{bdstate}
\ee
The fluctuation amplitude in this state in each fixed momentum channel is controlled by the r.m.s. value defined by 
$\delta \phi_k = || \phi_k ||$, where $\phi_k$ is the physical field with covariant canonical normalizations in, e.g. Eq. (\ref{unigr}). 
Here the fluctuation ``norm" in the weak coupling limit is given by the \underbar{\it renormalized} momentum space 
two-point function in the coincident limit (see, e.g. \cite{Linde:2005ht}), since non-Gaussianities are negligible when the couplings are small. Using the conformal time normalized field variable
$\varphi = a \phi$, this gives
\be
|| \phi_k ||^2_{renormalized} \, \delta^{(3)}(\vec k - \vec q) = \frac{k^3}{2\pi^2 a^2} \langle 0(\bar \tau)| \varphi_k(\tau) \varphi^\dagger_{q}(\tau) | 0(\bar \tau) \rangle 
|_{renormalized} \, ,
\label{twopt}
\ee
where we take $\bar \tau$ to be an instant during preinflation era after the initial inhomogeneities and anisotropies
are diluted away to below the signal generated by subsequent evolution. This means that there were $\sim {\cal O}(10)$
efolds prior to $\bar \tau$ \cite{Kaloper:2018zgi}. Renormalization involves subtracting the flat-space limit -- i.e. the UV limit -- 
of $\langle 0(\bar \tau)| \varphi_k(\tau) \varphi^\dagger_{q}(\tau) | 0(\bar \tau) \rangle$ at every momentum $k$. As we will see
below, this is just the standard normal ordering of the harmonic oscillator modes, subtracting away the $``1/2"$ from the zero-point
fluctuations. Using the formulas for the Heisenberg operators $b, b^\dagger$ a straightforward calculation yields
\be
|| \phi_k||^2_{bare} = \frac{k^3}{2\pi^2 a^2} \frac{ |{\cal M}_k(\tau,\bar \tau) + {\cal N}^*_k(\tau,\bar \tau)|^2 }{2k} \, .
\label{intensity}
\ee

How this fluctuation intensity imprints power to perturbations of the background now depends on the momentum $k$. To complete the calculation, and
also develop some intuition 
about it, let's look at some simple limits. Consider first the flat space limit, where $a = a_0 = {\rm const}.$, and $u_k = e^{-ik\tau}/\sqrt{2k}$ exactly. 
Thus ${\cal F}_k \equiv e^{-ik\tau}$, ${\cal G}_k \equiv 0$, and so ${\cal M}_k (\tau,\bar \tau) = e^{-ik(\tau - \bar \tau)}$, 
${\cal N}_k(\tau,\bar \tau) = 0$, and so 
\be
|| \phi ||^2_{flat~space} = \frac{k^2}{4\pi^2 a^2} = \frac{p^2}{4\pi^2} \, .
\label{radimode}
\ee
Here we used $p = k/a$ which is the physical, rather than comoving momentum.  Note that this formula will be valid at all momenta up to 
the cutoff $\Lambda_{UV}$, which might be as high as $M_{Pl}$, in principle, but needs to be at least higher than $H$ so we have any chance
of describing the locally flat limit to begin with. 
Next imagine 
that the universe is radiation dominated and take this epoch to be arbitrarily long,
with no intervening stages. In this case the wavefunction {\it in conformal time} is still $u_k = e^{-ik\tau}/\sqrt{2k}$. The 
$S$-matrices are simply unity (as there are
no interfaces). Hence we still find ${\cal F}_k = e^{-ik\tau}$, ${\cal G}_k = 0$, ${\cal M}_k(\tau,\bar \tau) = e^{-ik(\tau - \bar \tau)}$ and 
${\cal N}_k(\tau,\bar \tau) = 0$. As a result, the formula for $ || \phi ||^2_{bare~radiation}$ remains exactly the same 
as in the flat space limit (\ref{radimode}). 

We see that in both cases the r.m.s. fluctuation is directly proportional to momentum for any momentum $p$. This follows from the
fact that the ensuing scale invariance of the system implies that the only intrinsic scale for radiation is the momentum.  
In fact this is just the virial theorem for harmonic oscillators: an amplitude of a fluctuation on the average is given
by its inverse wavelength. If these fluctuations were on shell, they would dilute forever during
cosmic evolution, according to $|| \phi ||^2 \propto p_0^2 (a_0/a)^2$. But since these are vacuum fluctuations, these are nothing else but the zero-point fluctuations arising due to the
Heisenberg's uncertainty principle. These expectations can in fact be set to zero
by normal ordering -- or in other words, the term (\ref{radimode}) is precisely the counterterm which needs to be subtracted 
from a general $|| \phi ||^2_{bare}$ to obtain $|| \phi ||^2_{renormalized}$. 

Indeed, consider now $ || \phi ||^2_{bare}$ during long inflation, approximated by a single patch of spatially flat de Sitter chart. In
this case, the initial wavefunction $u_k$ of Eq. (\ref{bcsin}) is in fact exact, since the $S$-matrices are unity, and the direct substitution into
formulas above gives ${\cal M}_k(\tau,\bar \tau) + {\cal N}^*_k(\tau,\bar \tau)|_{\bar \tau \rightarrow -\infty} = \sqrt{2k} u_k(\tau)$. Therefore
\be
|| \phi_k||^2_{bare~dS} = \frac{k^3}{2\pi^2 a^2} |u_k(\tau)|^2 = \frac{k^3}{2\pi^2 a^2} \frac{k^2 + H^2 a^2}{2k^3} = \frac{1}{4\pi^2} \Bigl(p^2 + H^2 \Bigr) \, .
\label{intensityds}
\ee
where we have again written the final result in terms of the physical momentum $p = k/a$. So the two-point function has the
same UV behavior is it did in flat space, but due to the finite curvature radius of de Sitter it receives an IR correction $\sim H^2$. 
As is well known, de Sitter space behaves like a finite box. Now, subtracting (\ref{radimode}) from (\ref{intensityds}) finally yields the
renormalized two-point function in de Sitter,
\be
|| \phi_k ||^2_{renormalized~dS} = || \phi_k ||^2_{bare~dS} - || \phi_k ||^2_{flat~space} = \Bigl( \frac{H}{2\pi} \Bigr)^2 \, .
\label{dsfluctren}
\ee
This is the classic result of QFT in de Sitter space, equating the amplitude of a light quantum fluctuation with the 
Gibbons-Hawking temperature $T_{GH} = H/2\pi$ \cite{gibbhawk}. Clearly, this is an exact result in de Sitter space, valid for all $k$. 

Let us now turn to a more general evolution where de Sitter is interrupted by other epochs instead of being eternal. Specifically, we will work with  radiation epochs as the intervening eras, as in
our rollercoaster proposal. Let us first consider a single transition, from pre-rollercoaster accelerated stage to radiation. 
Start with a fluctuation in an epoch of accelerated expansion, required to prepare the
BD state. For formal reasons -- so as to technically eliminate any concern about the initial remnants dating to the prehistory of rollercoaster -- stretch the duration of this era from arbitrarily
far in the past to some long but finite time $\tau_0$ in the future. In terms of our parameterization in conformal time, keeping the
end of pre-rollercoaster fixed at $\tau_0$, imagine taking the beginning at $\bar \tau$ to be at a large negative value. In this case
the initial wavefunction for each mode is still (\ref{bcsin}), 
$u_k(\tau) = (1 + \frac{i}{k(1+\tau_0- \tau)}) e^{-ik(\tau-\tau_0)}/{\sqrt{2k}}$. After the transition to radiation era, 
the initial wavefunction (\ref{bcsin}) evolves to (\ref{radw1}), for times $\tau > \tau_0$: $u_k^{(1)} =\frac{1}{\sqrt{2k}} \Bigl[ \Bigl(1 + \frac{i}{k} - \frac{1}{2k^2}\Bigr) e^{-ik(\tau-\tau_0)} 
+ \frac{1}{2k^2} e^{ik(\tau-\tau_0)} \Bigr]$. It is now straightforward to use these to construct functions ${\cal F}$, ${\cal G}$, ${\cal M}$ and
${\cal N}$ for various combinations of initial and final times and for a given comoving momentum $k$.

To understand the implications of the transition for cosmological 
observables, we can focus to the leading order in wavelengths which are short and long, respectively, relative to the Hubble horizon length.
We need to however also consider their time evolution since if the background is not eternal de Sitter, the wavefunction and wavelength evolution
in time will eradicate the IR cutoff $\sim H_{dS}$. 
We start by looking at the the limits $k \gg 1$ and $k \ll 1$ at the time $\tau \la \tau_0$. The modes with comoving momenta 
$k \gg 1$ will not have managed to leave the horizon by this time -- they are very short wavelength. In contrast, the
modes $k \ll 1$ are long wavelength modes whose wavelength will have stretched beyond the horizon before $\tau = \tau_0$ -- i.e., before the epoch of accelerated expansion ended. So in these two limits,
as $\tau \sim \tau_0$, when $k \gg 1$, 
${\cal F}_k \rightarrow e^{ik(\tau - \tau_0)}$, ${\cal G}_k \rightarrow 0$. When $k\ll 1$ we have to be a bit more careful and keep subleading terms as well, since the leading terms $\sim 1/k^2$ cancel. 
So ${\cal F}_k \rightarrow (1+\frac{i}{k(1+\tau_0-\tau)} - \frac{1}{2k^2(1+\tau_0 - \tau)^2}) e^{-ik(\tau-\tau_0)}$, 
${\cal G}_k  \rightarrow \frac{1}{2k^2(1+\tau_0 - \tau)^2} e^{-ik(\tau-\tau_0)}$. 
At early times when $\bar \tau \rightarrow - \infty$, any $k$ mode is subhorizon for a sufficiently large $|\bar \tau|$.
So $\bar {\cal F}_k \rightarrow e^{ik(\bar \tau - \tau_0)}$, $\bar {\cal G}_k \rightarrow 0$ for both $k \gg 1$ and $k \ll 1$. 
Moreover, inflation will end at $\tau_0$ and radiation era will set in afterwards. 

Using these limits we can now construct
$|{\cal M}_k(\tau,\bar \tau) + {\cal N}^*_k(\tau,\bar \tau)|^2$ for both cases. The algebra is straightforward but tedious; 
we note that in the limit $k \ll 1$ but $\bar \tau \rightarrow -\infty$ we have 
\be
{\cal M}_k(\tau,\bar \tau) + {\cal N}^*_k(\tau,\bar \tau)  \rightarrow 
({\cal F}_k(\tau) + {\cal G}_k(\tau)) {\cal F}^*_k(\bar \tau) = \frac{ie^{-ik(\tau - \tau_0)}}{k(1+\tau_0 - \tau)}  = \frac{iHa(\tau)}{k} 
e^{-ik(\tau - \tau_0)} \, . 
\ee
Substituting this into (\ref{twopt}) yields $|| \phi_k||^2_{bare} = (H/2\pi)^2$. In the limit $k \gg 1$, the expectation value $||\phi_k||^2_{bare}$ is obviously the same as in (\ref{radimode}), 
given the asymptotic formulas for ${\cal F}, {\cal G}$. The additive renormalization -- i.e. the subtraction of the flat space limit of Eq. (\ref{radimode}) will therefore set it to zero, to the leading order in the $1/k$ expansion. 

Finally we should consider what happens with these modes when $\tau \gg \tau_0$, late into the radiation era. In this case we
should use the wavefunction (\ref{radw1}) in the construction of  ${\cal F}$, ${\cal G}$, ${\cal M}$ and
${\cal N}$ at late times. A straightforward albeit tedious calculation shows that the limiting form of the 
bare two-point function in this case yields $|| \phi_k||^2_{bare~dS} \rightarrow \frac{1}{4\pi^2} (p^2 + H^2/a^2 + \ldots)$ for
$p \ga H$, reproducing 
the flat space UV behavior but also showing that the IR cutoff contributions from past de Sitter chart are a transient effect, freezing
out superhorizon modes. But as those come back into the horizon, their power diminishes to zero as a power of environmental temperature
of radiation. The point which concerns us as it impacts the generation of perturbations in rollercoaster cosmology can be recapitulated 
as follows:
\be
|| \phi_k ||^2_{pre-rollercoaster} \rightarrow \begin{cases}
0 + \ldots \, , 
~~~~~~~~~~~~  {\rm for~~} k \gg 1  ~ {\rm at~very~late~conformal~times} \, ;\\
\\
\Bigl(\frac{H}{2\pi} \Bigr)^2 + \ldots \, ,  ~
~~~~  {\rm ~for~~} k \ll 1 ~ {\rm as~long~as}~ k < a H\, .
 \end{cases}
\label{inflation} 
\ee
The $\ldots$ stand for the higher order corrections induced by the interface between the accelerating phase and the radiation stage. 
This extends the result in de Sitter space, where the fluctuations in a 
de Sitter space behave like free harmonic oscillators at wavelengths short compared to the horizon $1/H$, which can be subtracted away
while leaving only the IR correction. A real inflation behaves as a hot, but leaky, can. We also stress that the interface(s) will induce mode `ringing'
as a function of momentum when the transitions are sudden, because of the phase rearrangements during the transition as a function
of the momentum. The ringing will be suppressed when the transitions are smooth, but the behavior is merely
a transient effect in the 
course of cosmic evolution. 

Note that in reality, instead of taking the limits $k \gg 1$ and $k \ll 1$ we
would fix a value of $k$ at some time $\bar \tau \la \tau \ll \tau_0$, evaluate $||\phi_k||^2$ and evolve it in time using 
(\ref{intensity}). In local causal QFT we would take initial modes to be causally correlated, i.e. with initial physical momenta 
$p = k/a(\tau) \gg H$. If the time evolution didn't grow the wavelength up to $H^{-1}$ by $\tau = \tau_0$, the mode would remain a quantum
fluctuation, which would be cancelled out by the subtraction of the flat space result, $||\phi_k||^2_{renormalized} \rightarrow 0$. 
If on the other hand the time evolution stretched the wavelength to superhorizon
scales by $\tau = \tau_0$, the mode would classicalize, with $||\phi_k||^2_{renormalized} \rightarrow (H/2\pi)^2$, which would stay approximately constant as long as $p < H$. What happens is a consequence of causality for superhorizon wavelengths. It is easy to show that the phase of each mode, which depends on time, 
$\propto \pm k \tau$, can be rewritten as $\pm k\tau = \pm k/{\cal H} + \delta = \pm p/H + \delta$ up 
to a constant phase shift $\delta$, where ${\cal H} = a'/a$
is the conformal Hubble parameter, $H$ the physical Hubble parameter and $p = k/a$ the physical momentum. 
This means that a mode which is superhorizon, $k/{\cal H} \ll 1$ has approximately constant phase, while a
subhorizon mode $k/{\cal H} \gg 1$ oscillates a lot. As a result, for modes $k/{\cal H} \ll 1$ amplitude remains nearly
constant, while the oscillating modes $k/{\cal H} \gg 1$ virialize, with amplitude $\propto k/a$. The former modes 
``freeze out", remaining a shadow of the burst of cosmic acceleration which put them on shell and expelled it out of 
the horizon, while the latter dilute away. 

As we already stressed, the formalism we set up above directly applies to a general rollercoaster cosmology as well. 
The only `generalization' of the mode functions is their \underbar{\it linkages} at the interfaces of various epochs according
to the matching conditions in Sec. 5.1. The key formula there is Eq. (\ref{matrixbcsoln}), which fixes the coefficients $A_k^{(n+1)}, B_k^{(n+1)}$
unitarily, using the propagation $S^{(j+1),(j)}$-matrices that encode the evolution of a mode during the entire epoch bounded by 
$j^{\rm th}$ and $(j+1)^{\rm st}$ interfaces. In each such region, the modes are either the Hankel functions in epochs of 
accelerated expansion, or flat-space free waves (in conformal time) in radiation era. 

So 
in a nutshell, the $S^{(j+1),(j)}$-matrices which link the modes up between different eras 
are analogous to the sequence of operators which interpolate the
M\"oller wave operators in QFT between finite times, without taking the limit of infinite past or future. 
The formula (\ref{intensity}) clearly remains valid, since 
it is local in time, which therefore means that we are interested into the limiting form for $|| \phi_k ||^2$ late into the rollercoaster evolution. 
The only difference between this case and inflation is that the evolution functions ${\cal M}_k(\tau,\bar \tau)$ and 
${\cal N}_k(\tau,\bar \tau)$, propagating the initial mode from time $\bar \tau$ to $\tau$, include the $S^{(j+1),(j)}$-matrix elements 
in addition to the smooth time evolution in the epoch where $|| \phi_k ||^2$ is to be calculated. 

Nevertheless, in the limiting cases of interest to us the 
formula (\ref{inflation}) remains perfectly valid {\it to the leading order}. Indeed, imagine a rollercoaster model with several epochs of
intermittent acceleration separated by radiation. As long as the accelerating eras are longer than the radiation eras separating them, and a mode's wavelength grows longer than the horizon during any era of acceleration, this mode will `freeze out': its wavelength will become
larger than the instantaneous value of  $1/H$, and the amplitude will become nearly constant. When the subsequent era of radiation is short,
the mode's wavelength will shrink relative to the growing horizon, but there just won't be enough time for the mode to `thaw': the
fluctuation will remain superhorizon, and the mode's amplitude will remain approximately constant. This will continue in the subsequent 
stages, with each later stage of accelerated expansion `freezing' this mode ever more: the corrections to the leading order
term $\sim \frac{1}{k^2 \Delta \hat \tau^2}$ in the matrix elements of $S^{(j+1),(j)}$-matrices will be suppressed more and more. 
Note that the amplitude of the frozen mode will be given by the instantaneous value of $1/H$ at its horizon crossing: the modes which froze 
during the $j^{\rm th}$ era of acceleration will have intensity $\propto H_j^2$, the modes which froze during the $(j+1)^{\rm st}$ era 
will have intensity $\propto H_{j+1}^2$, and so on. In other words, the IR distortion of a mode's two-point function is set by the
curvature at the time the mode freezes out. Simply put, this follows from the locality of the mode function. 

Conversely, a mode which does not leave the horizon during any epoch of acceleration will have its leading order 
amplitude $\propto k^2$ cancelled by the subtraction of the flat space value. Further, its IR correction due to the cosmic curvature
will be suppressed by its short wavelength, and so to leading order this mode's two point function will asymptotically vanish 
at late times, as $\sim H^2/a^2$. At a finite time, however, the mode will appear to have a small nonzero contribution controlled
by the early cutoff; yet as the universe expands this term will diminish. We will therefore largely ignore this behavior below. 

The behavior of the modes which leave the horizon during later stages of rollercoaster will interpolate between these limits. 
A mode which becomes superhorizon during an intermediate acceleration epoch will receive $\sim (H/2\pi)^2$ contribution to its two-point
function, with $H$ set by the curvature during this epoch.
A mode that becomes just barely superhorizon very close to the end of an accelerating era, and so it reenters quickly during a subsequent radiation epoch, will not be frozen and its amplitude – barring any subsequent epochs of acceleration – will naturally decrease in time. 
This can be interrupted by subsequent stages of accelerating expansion, and new IR corrections which these induce. 

As a result, the amplitude of the modes as a function of $k$ 
will oscillate in momentum space when the transitions are sudden.
Again, this is because of the phase rearrangements during sudden transitions; the `ringing' would smooth
out for continuous transitions from acceleration to radiation and vice versa. The total spectrum of the modes formed throughout the evolution will therefore have a shape in momentum space resembling a stairway in the sky. 
We will see this clearly in the subsequent sections, where we will present the results
of numerical integration of the mode evolution through example rollercoasters, demonstrating explicitly these features. We also note 
that some of the earlier analyses of power spectra in double inflation \cite{polstar,misao} display special examples of this behavior. 

As we noted, we will see the corrections to the leading order behavior in (\ref{intensity}) clearly. 
When inflation is very long, or the period between two epochs of inflation is long, these corrections
would be small. However in our case, with shorter epochs of accelerated expansion these corrections, generated by quantum effects
such as sudden transitions between the quantum states at the interface of  
two epochs late into the rollercoaster (see, e.g. \cite{Kaloper:2003nv} for an example), while small, may 
be considerably larger at least in some specific scenarios. These corrections could encode indications of the 
quantum origin of the fluctuations, and may thus be of interest to understand better. Such corrections in the scalar spectrum would
be too short wavelength to avoid contaminations by subsequent cosmological dynamics. However perhaps they could
remain conserved in the short wavelength primordial gravity waves, if those ever become a subject of extensive experimental exploration. 
We will relegate this question for future work\footnote{For some
approaches to formulate this question see eg \cite{martjven,alb}.}.

\subsection{Power to the Curvature}

The wave power of the fluctuations, due to the infrared cutoff induced by the freezeout of wave modes in a locally de Sitter chart describing
an accelerating stage of the rollercoaster, transfers to the curvature perturbations in the standard fashion. For the tensor modes, the individual 
helicities of the metric fluctuations are literally the scalars discussed above. So to leading order, the tensor power is just
\be
P_{\tt T} = \frac{8 || h_k ||^2_{renormalized}}{M^2_{Pl}} = 
\frac{8 H^2_j}{(2\pi)^2 M^2_{Pl}} \, , 
  \, ~~~~~~~~~~ {\rm ~for~~} k \le H_j \, ,
\label{tensorpower}
\ee
where $h$ is the canonically normalized tensor helicity, and $H_j$ the Hubble parameter of the accelerating epoch during which the mode $k$ froze out.
In the numerator, one factor of two counts the two tensor polarizations, one comes from the definition of tensor power as the square of $h_{ij}$, and one comes from squaring the canonical normalization.
Of course, the precise dependence of the power on the momentum is more complicated, as we will see below
using explicit numerical calculations. Note that this formula as it stands is valid only for superhorizon modes. Modes which return to the
horizon, or never leave it, have their power diluted away by cosmic expansion. 

The power in the scalar perturbations is more complicated. In particular, there is a lot of model dependence on the specifics of the
model describing the scalar field which dominates during a particular inflationary epoch. As we noted in Sec. 4, however, we can treat each
separate stage of acceleration as a short inflation to the leading order. Therefore, in each accelerating epoch, the scalar power transferred to the curvature is
\be
P_{\tt S} = \Bigl( \frac{H_j}{\dot \phi_j} \Bigr)^2 || \phi_k ||^2_{renormalized} = \Bigl(\frac{H_j^2}{2\pi \dot \phi}\Bigr)^2  \, , ~~~~~~~~~  \hfill  {\rm ~for~~} k \le H_j \, ,
\label{scalarpower}
\ee
where $\dot \phi_j$ is the proper time derivative of the scalar dominating the cosmological energy density during this epoch. Here
again $H_j$ is the Hubble parameter during the epoch where the scalar froze out. During the intervening
radiation eras, the scalar power would change asymptotically toward zero, albeit, again, the precise expression is more complicated because of 
the model dependence of the subsequent chain of evolution. Concretely, the prefactors $\frac{H_j}{\dot \phi_j}$ in general not even involve
the same field, and therefore can only be defined as piecewise smooth functions of the universal cosmic time $t$. Nevertheless, the universality of gravity guarantees that during a stage of accelerated expansion the dominant field's fluctuations will be imprinted into the geometry as curvature perturbations.

We will discuss what the model dependent factors $H/\dot \phi$ could be like during individual stages of acceleration in some detail in the following section.
We will also present some phenomenological requirements which a successful model needs to satisfy, as well as provide a simple, heuristic example, which might do. 
As a consequence, in the explicit numerical examples which we will present in Sec. 6, we will only focus on the tensor power,
which depends mainly on the sequence of geometric phases of the rollercoaster, and is less sensitive to the details of scalar dynamics which drives them. 

We also stress again that we will work under the assumption that the scalars of the rollercoaster are distinct individual fields rather
than different phases of the same field, which experiences a complicated potential. This is mainly for the sake of simplicity, and benefits from the fact
that as long as we devise models that fit the currently observed scales on the sky, we can ignore the dynamics controlling the transitions
between different stages of the rollercoaster. Clearly, this dynamics is very interesting for the determination of the precise form of predictions. 
However to study them one does need precise models. This challenge needs to be addressed in future work.

\section{How to Build a Rollercoaster}

Since our rollercoaster scenario involves periods of accelerated expansion interrupted by radiation epochs, a simple realization 
to look for may involve a QFT involving multiple inflatons which decay rapidly into radiation. Models of inflation and the processes 
of reheating have been studied extensively, and one might feel that finding realizations which support rollercoaster evolution is 
easy. 

A cursory look suffices to show things are not as easy as they might appear. Consider for example the old `gold standard' of 
inflation\footnote{Albeit somewhat `tarnished' thanks to the pressure from the data.}, based on field theories with a power law potential, $V \sim \lambda \phi^p/p$ \cite{lindechaot}.
One might wonder if simply taking a sequence of such models, which couple to fields whose mass gap depends on the inflaton {\it vev}, might suffice.
As long as $\phi \ga M_{Pl}$, the power law potential is flat enough to ensure
that the field kinetic energy is subleading to its potential, and that the energy density approximates vacuum energy. Then,
with some fine tuning, if the mass of the intended decay products becomes small when the inflaton nears the end of the 
slow roll regime, the inflaton decay process would be enhanced to quickly damp the inflaton's motion.

The problem with this idea is that the requisite period of inflation should be short. Instead of 50 -- 60 efolds we now seek ${\cal O}(15)$.
Thus if we imagine that the inflaton starts at a $\phi_i \ga M_{Pl}$ and that the decay should occur near $\phi_f \sim 0$ where the field
finds its minimum, the standard formulas (see e.g. \cite{lindechaot,kalshe}) yield the relation between the spectral index of the scalar perturbations $n_S$
and the number of efolds ${\cal N}$ before the end, where the perturbations are frozen,
\be
n_S = 1 - \frac{q}{{\cal N}} \, ,
\label{nscho}
\ee
where for power law potentials $q = (p+2)/2$. 
Demanding that $n_S \sim 0.96$ to fit the data at ${\cal N} \sim 10$ efolds before this stage of acceleration ends yields $p \simeq -1.2$.
This immediately rules out many of the common polynomial potentials in one fell swoop unless we require that the decay products
become extremely light at some value $\phi > M_{Pl}$, ending inflation by a catastrophic decay of the inflaton at values $\gg M_{Pl}$. 
This may look fine in EFT, but the UV completion of such models, requiring a detailed
realizations at what seems to be super-Planckian scales, may be challenging \cite{vafa1,vafa2}. The so called swampland arguments 
may be conjectural \cite{wgc,oovafa}, but the specter of physics above the cutoff looms over this approach nonetheless, and special care is required. 

Note the crucial influence which the changed duration of the stage of inflation during which CMB anisotropies are generated has on the emergence of this obstruction.
For example, in the standard approaches to testing for inflation by using the observational signatures, it is {\it always} assumed that the 50 -- 60 efolds required to solve the problems with selecting the background geometry must happen in one go.
Under this assumption one works hard to develop smooth models of long inflation, where by necessity the CMB anisotropies are imprinted during the first 10 of the last 50 -- 60 efolds by the quantum fluctuations of the same field which dominates expansion
all the way to the end. This necessitates using ${\cal N} \sim 50$ in the formula (\ref{nscho}), which then, fitting to the observation of 
$n_S \sim 0.96$, selects $q=2 + {\cal O}(\epsilon)$. A power law potential this points to is the quadratic, but the quadratic produces too
much tensor power, which is now ruled out. The only other model which gives $q=2+{\cal O}(\epsilon)$ is the Starobinsky model 
\cite{Starobinsky:1980te} which can be written down as a quartic theory of gravity, where the slightly more complicated form of the potential
leads to a significantly lower value of tensor power. Thus one often hears nowadays that the Starobinsky model is a model of inflation favored by the data. 

Also note that the spectral index changes with ${\cal N}$, albeit not very fast. The variation of $n_S$ with scale can be characterized by the
running function $\alpha = - dn_S/d{\cal N}$, which for models that predict (\ref{nscho}) yields
\be
\alpha =  -\frac{q}{{\cal N}^2} = \frac{n_S - 1}{\cal N} \, ,
\label{running}
\ee
Clearly, in long smooth inflation which fits the data, $n_S \simeq 0.965$, $\alpha$ is quite small, being at most 
$\alpha \simeq -0.0007$.
This makes it difficult to observe; the current bounds on the running are $\alpha \simeq -0.0041 \pm 0.0067$ \cite{planck18}.
 
However the statement that the CMB anisotropies are generated during the first 10 of the last 50 -- 60 efolds of a single stage inflation is obviously an \underbar{\it assumption}.
The fitting of the spectral index to $q = 2 +{\cal }(\epsilon)$, therefore, is really not forced on us by data, but by a combination of data and the assumption that inflation happened all in one go. 
One might think this is reasonable by invoking some argument of simplicity. One may also take
a different point of view, as for example espoused in \cite{oovafa,vafa1,vafa2}, which posits that very 
long smooth inflation is very difficult, and perhaps impossible to achieve. 

We will not delve on the merits and demerits of either philosophy here. Instead we take a pragmatic attitude: we do not know whether
the inflationary stage early on was a long smooth one, or that the evolution was rollercoaster-like, with shorter bursts of accelerated
expansion intermingled by radiation epochs, or even some other stages. Therefore, let's test this {\it observationally}.

The one immediate theoretical question then is to outline a set of models which may be singled out by an affirmative answer to 
whether a rollercoaster preceded the usual hot big bang. Those models would clearly need to be different from both 
the power law potentials and the Starobinsky
model, given that they must produce CMB anisotropies at a different stage of evolution, where the efolds clock points at
$\ga 15$, instead of 50 -- 60, before the end of the relevant stage.  

This clearly throws the model building door wide open, and involves a task of reevaluating the usual suspects for inflationary models
in a new light. One key criterion which those models must pass is to reproduce the correct answer for the spectral index of the
scale dependence of the CMB anisotropies. For standard inflation this yields a classification of various models as, for example,
presented in \cite{alibaba}. The important distinction in rollercoaster is that the pivot point where the spectral index is normalized to
observations is {\it not} at ${\cal N} \simeq 60$ efolds before the end of the accelerated epoch, but at a much lower value -- e.g. 
${\cal N} \simeq 15$. As we already implied, this completely changes the model selection. Indeed, as is obvious from Eq. (\ref{nscho}),
reducing ${\cal N}$ requires flattening the dependence of $n_S$ on it in order to keep its normalized CMB value at $n_S \simeq 0.96$.

\begin{figure}[thb]
    \centering
    \includegraphics[scale=1]{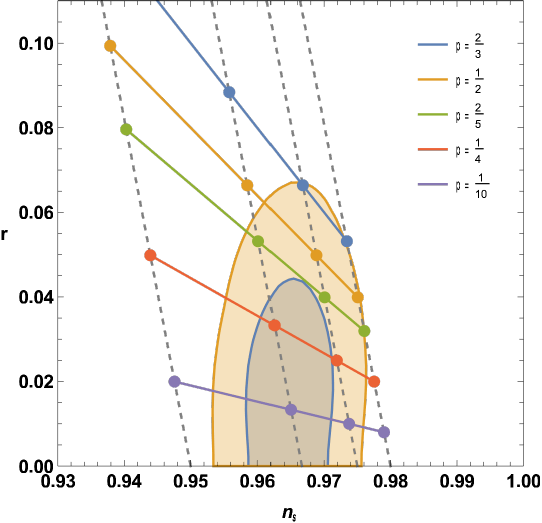}
    \caption{Tensor-to-scalar ratio versus spectral index for flattened power laws and the data. The blobs  correspond to, from left to right, 
    ${\cal N} = 20$, ${\cal N} = 30$, ${\cal N} = 40$ and ${\cal N} = 50$ efolds before the end of the accelerating evolution in this stage. 
    We use the data in \cite{bicep}. We see that for a fixed $p$ the data favor the band 
    of ${\cal N}$ of  $\sim$25 -- 40. Also, we observe that potentials flatter than $\sim \phi^{2/3}$ are not excluded, and that those
    flatter than $\sim \phi^{2/5}$ are in excellent agreement with the data at present, at lower values of ${\cal N}$.}
    \label{fig:potato1}
\end{figure}

We illustrate this by using power law potentials $V \sim \lambda \phi^p/p$, which however are very flat, with $0 < p < 1$. Such models
are known to arise in flattened monodromy models of inflation where the phenomenon of flattening comes from strong 
interaction effects \cite{eva1,eva2,evil,london,pisky} (see also \cite{seizing}). 
If we take $p < 1$ and ${\cal N} \sim$ 20 -- 40, as opposed to ${\cal N} \sim$ 50 -- 60, the fit of these models to the data improves considerably,
as shown in Fig. \ref{fig:potato1}. In this case, $q \simeq 1+ p/2$ is closer to unity than to $2$ -- a significant difference compared to 
what happens in a single long stage of inflation. Note also that the spectral running is also bigger for these models. Using (\ref{running}) we
find that $\alpha \simeq - 0.035/20 \sim - 0.00175$, which while still small may be more accessible to observations than in the case of long smooth inflation. 
It would be very interesting to check if a ladder of monodromies might provide for
a consistent rollercoaster evolution.

To illustrate further how the changed perspective alters the early universe cosmology, 
we will in fact focus on a special subclass of models where the flattening of $n_S({\cal N})$ is 
extremal: namely, we consider the cases where spectral index is completely independent of the efold clock value ${\cal N}$. The models which
yield this behavior are the well known theories with exponential potentials \cite{abbwi,luma}, which yield scale factors $a(t) \sim t^\zeta$, 
$\zeta \gg 1$. We imagine that during an epoch of acceleration of the rollercoaster
universe, the potential is well approximated by the exponential function,
\be
V(\phi) = V_0 e^{c \phi/M_{Pl}} \, ,
\label{expot}
\ee
where $c$ is a dimensionless constant, and $V_0$ is degenerate with the initial value of $\phi$. Interestingly such dilatonic potentials
have been invoked in the attempts to evade problems with swampland bounds \cite{vafa1,vafa2}, and are commonplace in the literature.
During this era, the well-known late time attractor solution \cite{abbwi,luma,liddlexp,kalol} is quickly reached, and it describes the cosmic evolution during $\phi$-domination: 
$a = a_0(t/t_0)^{2/c^2}$, $\phi = \phi_0 - \frac{2}{c} M_{Pl} \ln(t/t_0)$.
Taking $t_0$ as the instant when the attractor evolution starts to dominate, with initial value $\phi_0$, and introducing ${\cal N} = \ln(a(t)/a_0)$ as the number of efolds that transpired until time $t$, we find that the 
field variation is $\Delta \phi/M_{Pl} = - c {\cal N}$, and that the scalar power, tensor power, spectral index $n_S$ and the 
tensor-scalar ratio during this epoch are
\be
P_{\tt S} = P_{\tt S}(0) \Bigl(\frac{k}{k_0}\Bigr)^{-2c^2/(2-c^2)} \, , ~~~~~~~ P_{\tt T} = r \, P_{\tt S} \, , ~~~~~~~ n_S = 1- \frac{2c^2}{2-c^2} \, , ~~~~~~~ r = 8 c^2 \, ,
\ee
where $P_{\tt S}(0), k_0$ are the Planck normalization values, $P_{\tt S}(0) \simeq 2.1 \times 10^{-9}$ \cite{planck18}. These formulas
are totally independent of ${\cal N}$, which is only determined by the variation of $\phi$ in the field space,
$c{\cal N} \simeq \Delta \phi/M_{Pl}$. Note that as a consequence in these models the spectral running vanishes, $\alpha = 0$. 

\begin{figure}[thb]
    \centering
    \includegraphics[scale=1.0]{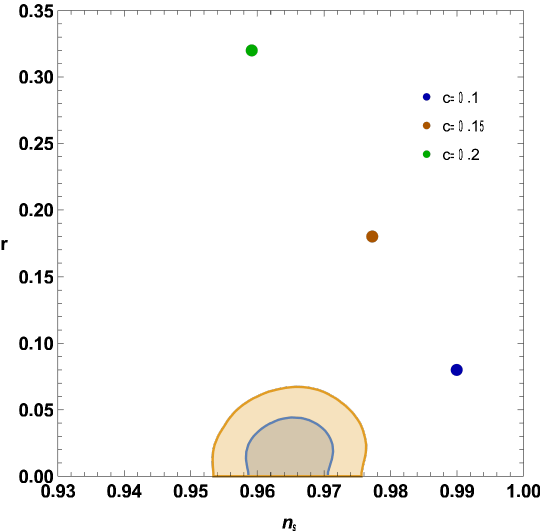}
    \caption{The tensor-to-scalar ratio versus the spectral index and the CMB data. Again, the data is from \cite{bicep}. 
    Although they are not excluded yet, even the single term exponentials are under pressure from the data. }
    \label{fig:potato2}
\end{figure}

The general trend is displayed in Fig. \ref{fig:potato2}, using most recent data from \cite{bicep}.
It is clear that pure exponential potentials are excluded by CMB data in plain vanilla $\Lambda$CDM late universe, although they can become viable again if the fit to $n_s$ changes because of different late-time cosmology~\cite{DAmico:2021fhz}. Their prediction cannot be modified by changing the number of efolds at which large-scale perturbations are generated.

We also note that for for values of $c \simeq 0.2$ the field space excursion
of $\phi$ is rather modest: the key is that we only need to ensure this epoch goes on for $\sim {\cal O}(15)$ efolds, no more. This gives 
$\Delta \phi/M_{Pl} \simeq {\cal O}(15) \times c \simeq {\cal O}(3)$. Note of course that the fits in Fig, (\ref{fig:potato2}), being completely independent
of ${\cal N}$, could be taken to represent long smooth inflation as well. However, in stark contrast to short rollercoaster epochs, if the exponential 
potential is to represent a long smooth period of inflation, the field variation must be significantly larger to support
$60$ efolds. Indeed, we'd need $\Delta \phi/M_{Pl} \simeq {\cal O}(60) \times c \simeq {\cal O}(10)$, almost an order of magnitude larger field space for $\phi$. It is unclear at the moment if such large field variations for exponential potentials are realistic \cite{vafa1,vafa2}.

It would be interesting to see if a full rollercoaster could be built out of stages involving acceleration driven by exponential potentials.
Here we will merely outline what is needed in order to accomplish this. First of all the exponential potentials, once they set into the 
attractor mode, tend to dominate the evolution forever or remain subleading forever in FRW cosmologies with a fixed dominant component.
Here we need to switch the exponential potential (\ref{expot}) on and off. We can in fact imagine an EFT mechanism that can play role 
of the switch. Suppose that the potential is really $e^{c(\phi/M_{Pl}) \phi/M_{Pl}}$, where early on $c(\phi/M_{Pl}) \gg 1$. In this regime,
the field will roll very slowly and yet its energy will dilute very fast, since it scales as $e^{c(\phi/M_{Pl}) \phi/M_{Pl}}$. The field will be
subleading to other contents of the universe for a long time. As $\phi$ changes by ${\cal O}(1)$ a phase transition can occur,
as argued for example in \cite{vafa1,oovafa}, and the corrections to the dynamics -- for example from many fields which suddenly became
light \cite{reece,gpv,kalpre} -- can renormalize the EFT in the new vacuum yielding a new value of $c$ much smaller than unity. 
The reduction of value of $c$ will yield to slower dilution of energy density in $\phi$ and a stage of acceleration can kick in. The corrections might also
affect the fit to the data, perhaps improving the $r$ vs. $n_S$ compatibility in Fig. \ref{fig:potato2}. 

Once the field traverses additional range, its couplings to more degrees of freedom in the theory may become large
and those degrees of freedom become light, as in examples considered in \cite{reece,gpv,kalpre}. The stage of accelerated expansion 
may end there since 1) the newly light degrees of freedom correct the dilaton potential and steepen it, causing a dilaton vacuum
shift and 2) the dilaton quickly decays into these degrees of freedom, dumping its energy into relativistic radiation, as in \cite{kalpre}. 
A cartoon of a theory might be a dilaton coupled to many fermions via terms like $\partial_\mu \phi \bar \psi \gamma^\mu \psi$, and with
fermion mass terms like $m (1-g_1 e^{g_2 \phi/M_{Pl}}) \bar \psi \psi$. In this case, many light fermions could renormalize the kinetic term via
$(\partial \phi)^2 \rightarrow Z (\partial \phi)^2$ as in \cite{pisky,seizing,kalpre}, while keeping the corrections to (\ref{expot}) at ${\cal O}(1)$ 
level. This would reduce $c_{eff}$ by a factor of $1/\sqrt{Z}$, which could be significant if many fermion modes are present.
A more complicated setup can result in a modified kinetic term, which reduces the speed of sound of fluctuations. This way, the value of $r$ is lower and exponential potentials could still satisfy CMB constraints~\cite{Unnikrishnan:2013vga}.
We plan to explore such setups in more detail in the future.

If this type of dynamics can be set up in microscopic theory, 
the rollercoaster is ready to cast off. We then need to provide additional light fields like $\phi$, to drive subsequent stages
of accelerated expansion. Those might be fields similar to the dilaton, albeit that is not necessary. Here we will not delve into a
detailed explorations of these stages, since they are not very constrained by the data at present. As long as they are longer
than the intervening periods of radiation, and they produce a total of at least 50 -- 60 efolds of accelerated expansion before,
say, QCD phase transition and nucleosynthesis, they will do. 

\section{Rollercoaster Examples}

We will consider several cases
here, starting with the simplest case of all: 
\begin{itemize}
\item
The case where a stage of preinflation is separated from one more stage of
cosmic acceleration by a stage of radiation domination, and involving a final stage of radiation domination; all in all,
a short rollercoaster of four stages;
\item
The case with three stages of accelerated expansion, including preinflation, separated by two radiation stages and ending with the third
radiation stage, all of equal duration; this case is phenomenologically problematic because the equal duration 
of the acceleration and radiation epoch allows for too much mode evolution even for small $k$; the modes do 
not freeze out completely during the full evolution since they keep coming back into the horizon too soon; 
\item
The case with three stages of acceleration, where the zeroth is preinflation, and the other two each yield 25 
efolds of accelerated expansion, which are separated by mutually equal but shorter stages of radiation domination, 
each lasting 3 efolds, in between, and finally ending with a long stage of radiation domination.
\end{itemize}
The first case is quite simple. In the subsection below we will write the solutions in the analytic form in full detail, in order to
illustrate the procedure for extracting the physical observables and make the physical features of the evolution
abundantly clear. In the next subsection, we will consider the two later cases. We will solve for the wavefunctions iteratively, using the 
discussion of Sec. 5.2, and compute the power in the wave modes numerically, plotting the results. 
It is possible to write the wavefunctions 
analytically -- after all we found the solution, which is given in Eq. (\ref{matrixbcsoln}) -- but the equations are very cumbersome. Instead we will present the specific form of the solutions 
graphically in order to compare them with the predictions 
from standard long and uninterrupted inflation. 

\subsection{Cosmologia con Quattro Stagioni}

\begin{figure}[thb]
    \centering
    \includegraphics[scale=0.9]{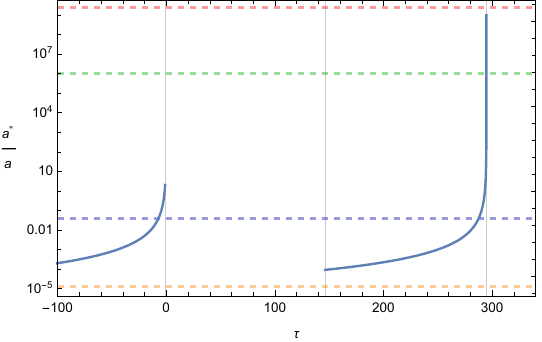}
    \caption{Effective potential $a''/a$ for the Mukhanov-Sasaki equation~\eqref{eq:maineq}. The horizontal dashed lines correspond to the fixed comoving wavenumbers $k$ indicating periods when they are subhorizon $k > {\cal H}$
    and superhorizon $k < {\cal H}$.}
    \label{fig:potential}
\end{figure}

Here we consider the case where a long stage of preinflation is interrupted by an epoch of radiation, which is then replaced by one more 
stage of cosmic acceleration of finite duration, that ends in radiation. The final transition is analogous to the reheating stage at the end
of the usual long smooth inflation, and represents the Hot Big Bang, signaling the onset of the Standard Model of cosmology. 
We are interested in the evolution of the perturbations evolving through the various epochs. The wavefunction is the solution of the Mukhanov-Sasaki 
equation (\ref{eq:maineq}), and its evolution is governed by the effective potential $\nu = a''/a$, which is depicted in Fig. \ref{fig:potential}. 
\begin{figure}[thb]
    \centering
    \includegraphics[scale=0.9]{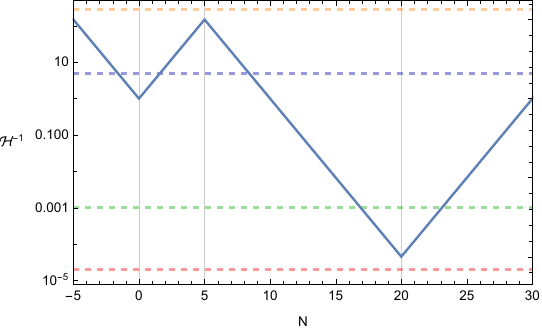}
    \caption{Comoving wavelength $1/k$ versus conformal horizon $1/{\cal H} = a/a'$. The horizontal lines correspond to the fixed 
    comoving wavelengths $1/k$ indicating periods when they are superhorizon $1/k > 1/{\cal H}$ and subhorizon $1/k < 1/{\cal H}$
    as a function of time ${\cal N}(t)$. The fixed values of $k$ are identical to those in Fig. \ref{fig:potential}. }
    \label{fig:vivaldi}
\end{figure}

The evolution of this wavefunction is controlled by the horizon crossing condition, $k = a H$, which in terms of the conformal `Hubble rate' is just
$k = a'/a$. Clearly, the prime denotes a conformal time derivative. During an epoch of acceleration, a mode with a sufficiently large $k$ is subhorizon,
$k > a'/a$, and therefore the wavefunction propagates through classically allowed regions of the potential $\nu$; as time goes on, the modes
may `leave' the horizon: $a'/a$ and $a''/a$ grow, and $u_k$ evolves into the classically prohibited region, where oscillations are replaced
by damping. Modes which stay superhorizon until the final reentry late during the last stage of radiation will retain their amplitude for a very long 
time. This is the source of scale invariance of the subsequent perturbations imprinted on the curvature. Some modes may reenter during the
intermediate epoch of radiation, and leave again during the second stage of acceleration. Their amplitude may be `bumped' during this stage,
eventually converging to the value of the physical Hubble parameter at the final transition to radiation. We depict the $1/k$ versus $a/a'$ 
behavior in Fig. (\ref{fig:vivaldi}). 

We can describe the evolution of the wavefunction in analytic form straightforwardly. 
The initial wavefunction, due to the long duration
of inflation which sets up the BD vacuum, is given in (\ref{bcsin}). At the interface of the end of preinflation and the subsequent finite era of
radiation, this wavefunction evolves into $u_k^{(1)}$ of Eq. (\ref{radw1}). 

At the end of the radiation era and the onset of the final stage of
acceleration, the wavefunction reverts back to the linear combination of Hankel functions, with the coefficients set by the smooth boundary
conditions at the interface. A straightforward application of the equations of Sec. 5.1 shows that the wavefunction during the second
stage of acceleration evolved to 
\begin{multline}
   u^{(2)}_k(\tau) = \frac{A_k^{(2)}}{\sqrt{2k}} \( 1 - \frac{i}{k (\tau - 2 \tau_1 + 2 \tau_0)} \) e^{- i k (\tau - 2 \tau_1 + 2 \tau_0)} + \\
    \frac{B_k^{(2)}}{\sqrt{2k}} \( 1 + \frac{i}{k (\tau - 2 \tau_1 + 2 \tau_0)} \) e^{i k (\tau - 2 \tau_1 + 2 \tau_0)} \, .
    \label{acc2u}
\end{multline}
To ensure the time $\tau$ changes continuously across the interface, the functional form (\ref{acc2u}) involves shifted time-dependent 
arguments. The boundary conditions relate the coefficients to those in (\ref{radw1}), which yields
\be
 A_k^{(2)} = -\frac{1}{4 k^4 \tau_0^2 x_1^2} + \frac{e^{-2 i k (\tau_1 - \tau_0)}}{4 k^4 \tau_0^2 x_1^2} \( 1 + 2 i k x_1 - 2 k^2 x_1^2 \) \( 1 + 2 i k \tau_0 - 2 k^2 \tau_0^2 \) \, ,
 \ee
 \be    B_k^{(2)} = \frac{e^{-2 i k \tau_0}}{4 k^4 \tau_0^2 x_1^2} \( 1 + 2 i k \tau_0 - 2 k^2 \tau_0^2 \)
    + \frac{e^{2 i k x_1}}{4 k^4 \tau_0^2 x_1^2} \(- 1 + 2 i k x_1 + 2 k^2 x_1^2 \) \, , 
    \label{coeffs2acc}
\ee
where for simplicity of notation we introduce $x_1 = \tau_1 - 2 \tau_0$.

When the second stage of acceleration ends, the wavefunction transmutes again into a (conformal time) plane wave, 
\begin{equation}
  u^{(3)}_k(\tau) = \frac{A_k^{(3)}}{\sqrt{2k}} e^{- i k \tau} +  \frac{B_k^{(3)}}{\sqrt{2k}} e^{i k \tau} \, .
\end{equation}
where the matching conditions at the interface at $\tau = \tau_2$ with $u^{(2)}_k$ 
give
\begin{eqnarray}
       A_k^{(3)} &=& -\frac{1}{8 k^6 \tau_0^2 x_1^2 x_2^2} \( 1 + 2 i k x_2 - 2 k^2 x_2^2 \) \( 1 + 2 i k x_1 - 2 k^2 x_1^2 \) \( 1 + 2 i k \tau_0 - 2 k^2 \tau_0^2 \) \nonumber \\ 
    &&+ \frac{1}{8 k^6 \tau_0^2 x_1^2 x_2^2} \bigg[ e^{2 i k (\tau_1 - \tau_0)} \( 1 + 2 i k x_2 - 2 k^2 x_2^2 \)
    + e^{2 i k (\tau_2 - \tau_0)} \( - 1 + 2 i k x_1 + 2 k^2 x_1^2 \) \nonumber \\
    &&+ e^{2 i k (\tau_2 - \tau_1)} \( 1 + 2 i k \tau_0 - 2 k^2 \tau_0^2 \) \bigg] \, , 
\end{eqnarray}
\begin{eqnarray}    
    B_k^{(3)} &=& \frac{e^{-2 i k (\tau_2 - \tau_1 + \tau_0})}{8 k^6 \tau_0^2 x_1^2 x_2^2} + 
    \frac{e^{-2 i k \tau_0}}{8 k^6 \tau_0^2 x_1^2 x_2^2} \(- 1 + 2 i k x_1 + 2 k^2 x_1^2 \) \(- 1 + 2 i k x_2 + 2 k^2 x_2^2 \) \nonumber \\
    &&+ \frac{e^{-2 i k \tau_1}}{8 k^6 \tau_0^2 x_1^2 x_2^2} \(- 1 + 2 i k x_2 + 2 k^2 x_2^2 \) 
    \( 1 + 2 i k \tau_0 - 2 k^2 \tau_0^2 \) \nonumber \\
    &&+ \frac{e^{-2 i k \tau_2}}{8 k^6 \tau_0^2 x_1^2 x_2^2} \(- 1 - 2 i k x_1 + 2 k^2 x_1^2 \) 
    \( -1 - 2 i k \tau_0 + 2 k^2 \tau_0^2 \)\, ,
\end{eqnarray}
where we again simplify the notation by using $x_1 = \tau_1 - 2 \tau_0$ and $x_2 = \tau_2 - 2 \tau_1 + 2 \tau_0$.

Having the exact solution is informative and satisfying, but quite cumbersome -- we wrote it formally in (\ref{matrixbcsoln}), but its explicit 
form is messy. In the end, since we are ultimately only interested in the
physical variable representing the (one mode) power spectrum (this is the same equation as (\ref{intensity})!), 
\begin{equation}
    {\cal P}(k) = \frac{k^3}{2\pi^2} \frac{|u_k(\tau)|^2}{a(\tau)^2} \, ,
\end{equation}
We are interested in the behavior of this variable as a function of $k$ and the conformal time. We focus on the tensor power because the
scalar power depends on details of the rollercoaster model, that controls the additional prefactor 
$({H}/{\dot \phi})^2 $ which mediates the
conversion of ${\cal P}$ to $P_{\tt S}$, as stated in Eq. (\ref{scalarpower}). In the end, as we saw in Sec. 7, all this amounts to is giving an
explicit functional form of the tensor-to-scalar ratio during each individual epoch of accelerated expansion and then connecting these together. 
Thus knowing ${\cal P}  \propto P_{\tt T}$ and $r$ will give us $P_{\tt S} = P_{\tt T}/r$ for any given model, albeit the full form will be a 
complicated and at-best a piecewise smooth function. We cannot expect otherwise if the stages of accelerated expansion are dominated by different
scalars. 

We will focus on ${\cal P}$ evaluated at a fixed conformal time $\tau$ which is the instant denoting the transition to the final radiation stage. This means
that we are defining ${\cal P}$ to be the initial condition for the primordial tensor power spectrum which will be processed in the late Universe.
For modes with very large wavelengths $k \ll 1$, the function ${\cal P}$ is quite insensitive to this choice. These modes are superhorizon,
and ${\cal P}$ remains very constant until late into radiation era, when $k \sim a'/a$ and the mode `thaws'. We will however 
also show what happens with the power function at a fixed $k$ as time goes on. For the purpose of presenting the 
results we will normalize ${\cal P}$ by dividing it with the power spectrum in the $k \ll 1$ limit which would have been produced by the 
stage of preinflation extended to last forever. This form factor function $F(k)$, 
\begin{equation}
    F(k) = \frac{{\cal P}(k)}{(H/2\pi  M_{pl})^2}\, ,
    \label{eq:ps}
\end{equation}
is normalized to unity as $k \rightarrow 0$, where $H$ is the almost constant Hubble parameter during the pre-rollercoaster stage of 
accelerated expansion.

\begin{figure}[thb]
    \centering
    \includegraphics[scale=0.9]{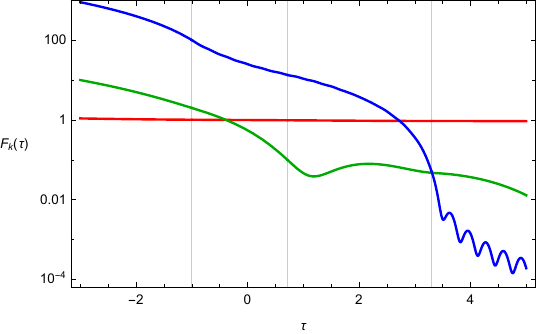}
    \caption{Time evolution of wave power at a fixed $k$. The background comprises preinflation
    and a 3-efold long epoch of acceleration, separated by an efold of radiation. We normalize 
    initial comoving momenta in units of physical $H$. Their values are $k=0.1$ (red), $1$ (green) and
    $10$ (blue).}
    \label{fig:Fktime}
\end{figure}

Let us now present the results of the numerical integration. 
First, we show, as a check of the discussion in Sec. (5.2), the evolution of $F(k)$ for a fixed comoving momentum $k$ as
function of time. The result is given in Fig. \ref{fig:Fktime}. We indeed see how the effects of early de Sitter-like stage on the
stored power in a frozen mode dissipate as the mode reenters: $F$ evolves from nearly a constant value to asymptotically zero in radiation, 
decaying at late times as a power law, as we observed in analytical considerations in Sec. 5.2.
This illustrates why the modes probed by the CMB must not thaw immediately after the era of acceleration when they are generated. Even if they refreeze again (see below) they will be
too affected by the evolution (see also \cite{misao,clifton,westy}). 

Next we present the power as a function of $k$, which shows what happens when a mode freezes. We see that
for very long wavelength modes, the form factor $F(k)$ tends to unity as $k \rightarrow 0$. These are modes 
which freeze out very early during the epoch of preinflation, and turn into a cosmic ``permafrost" since they spend a long time
out of causal contact. They will only reenter very late during the final radiation era. The shorter wavelength modes
may temporarily freeze out during the epoch of preinflation, but if this happens near the end, they could thaw during
intervening radiation era and evolve -- until they freeze out again. 

The power in these shorter modes will be bumped during 
the radiation era due to the phase shifts; since the wavelength is sufficiently short the relative phase will change by 
the next reentry into the barrier and alter the net amplitude relative to a long wavelength mode. In other words, the amplitude will
diminish due to the expansion of the background. 
However the change of the amplitude of these intermediate $k$ modes is bounded due to the second freezeout, which occurs after the finite era of acceleration starts, and the mode reenters the barrier. 
During this era the power would asymptotically approach $(\tilde H/2\pi)^2$, where
$\tilde H$ is the Hubble parameter during the second stage of acceleration, $\tilde H < H$. Thus the form factor would tend to a new value,
$F(k) \rightarrow (\tilde H/H)^2 < 1$ as $k$ gets larger. 

\begin{figure}[thb]
    \centering
    \includegraphics[width=0.49\textwidth]{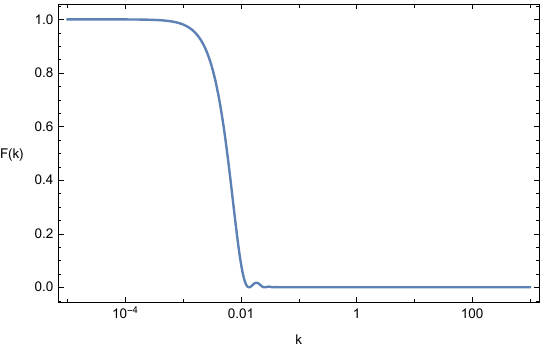}
        \includegraphics[width=0.49\textwidth]{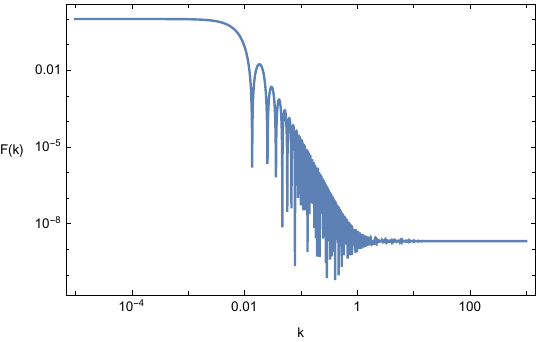}
    \caption{Power as a function of momentum; left panel, linear-log units, right panel log-log units. We always plot the $k$ axis in log units. The bumps in between
    the asymptotically constant regions represent the behavior of the modes which `thaw' at, or after the end of first stage of acceleration, and re-`freeze' during the second stage.}
    \label{fig:Fkk}
\end{figure}

This is seen clearly in Fig. \ref{fig:Fkk}, where we plot $F(k)$ as a function of $k$ both in linear and log-log coordinates.
The right panel in log-log coordinates is useful to resolve the behavior of modes which reenter during the intermediate radiation era and exit again during the last
epoch of radiation. We note that these results are in agreement with the analyses of \cite{slava1,polstar,misao,clifton}. Also, gravity waves which are generated during phase transitions well after rollercoaster or inflation behave like this \cite{johnco}. Indeed,
our principal reason for considering the simple four stage cosmos here is to match those results as a check. 

It is also quite clear 
that a) interruptions of accelerating stages by radiation stages only affect significantly the modes with momenta in the intermediate
range, which are not frozen throughout the radiation era; however if the radiation era is prolonged, more and 
more modes will be affected; and b) that if the second stage of acceleration is long
it could serve as inflation with the preceding era of radiation largely invisible to observers in a late universe for a long time.
However in the case b), if for any reason the observers could see close to the beginning of the last stage, they might 
discern variation of power as a function of scale due to the mode ringing. 
At the largest scales for such observers some modes would have more power, and some less, relative to the shorter scales. 

It is tempting 
to think if this can also be linked to the purported anomalies in the CMB at the largest scales on the sky (see e.g. \cite{punct}). 
However to check whether such 
an implication can be made requires a detailed model of a rollercoaster, where the adjacent epochs of accelerated expansion are
connected by a radiation era, with an explicit description of the dynamics of the fields which control different acceleration stages
and the transition between them. The plots we have examined are for the tensor power, and to infer what happens with
the scalar power, that might be anomalous, we need a precise analysis of $r$. 

\subsection{Oltre le Quattro Stagioni}

Now we turn to the cases with three stages of accelerated expansion, one of which is preinflation.
In between them there are two finite radiation stages while at the end there is the third stage of radiation.
In our first example, we take the intermediate acceleration and radiation 
stages to all be of the same duration, leaving the earliest stage of acceleration and the last stage of radiation to be of 
indeterminate length -- i.e., long. Our main interest in this case is to demonstrate that it is
phenomenologically problematic, because the spectrum of perturbations does not really have a scale invariant spectrum of perturbations because
the intermediate epochs of radiation are too long. Briefly, this happens because the modes which freeze out during acceleration stages
thaw too efficiently during the long intermediate radiation stages. The exponential suppression of the subleading terms in the mode functions
which occurs in long smooth inflation, or in the rollercoasters with short intermediate stages of radiation simply isn't there. 
\begin{figure}[ht]
    \centering
    \includegraphics[scale=0.9]{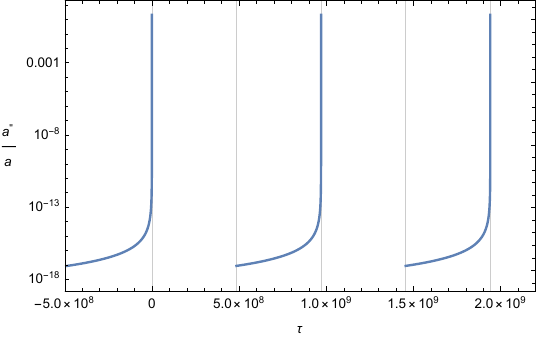}
    \caption{Effective potential $a''/a$ for the Mukhanov-Sasaki equation~\eqref{eq:maineq} for the case with 
    intermediate stages of equal duration.}
    \label{fig:EDpotential}
\end{figure}

Let us now demonstrate that visually, by displaying the plots of the numerical investigation of mode evolution in this case. 
In Fig. \ref{fig:EDpotential} we display the effective potential $a''/a$ for the perturbations, that clearly shows the three eras of
acceleration separated by two radiation epochs of the same length. The earliest acceleration epoch can be past-extended arbitrarily far.
Taking the initial wavefunction to be (\ref{bcsin}), justified by the pre-rollercoaster preparation of the BD state as the initial state of the evolution
we propagate the modes through the barriers.

\begin{figure}[hbt]
    \centering
    \includegraphics[width=0.49\textwidth]{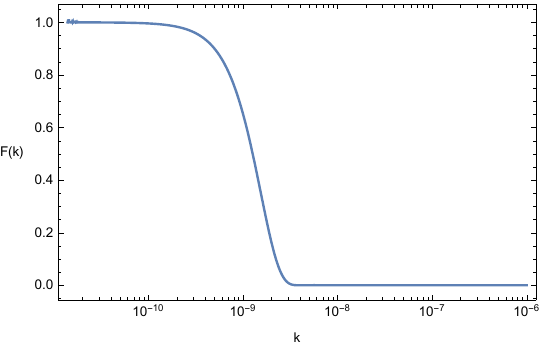}
        \includegraphics[width=0.49\textwidth]{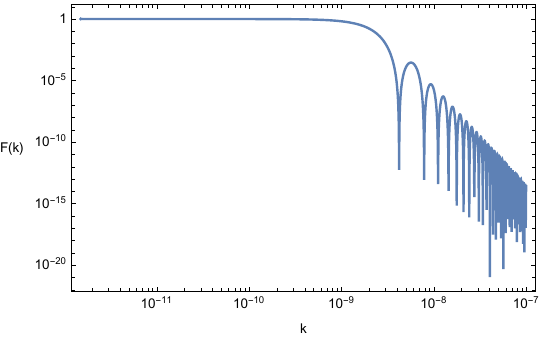}
    \caption{Power as a function of momentum; left panel, linear-log units, right panel log-log units. The intermediate stage of acceleration  
    is largely irrelevant for the mode evolution because it is too short compared by the stages of radiation before and after it.}
    \label{fig:EDFkk}
\end{figure}

The results for the power in the modes as a function of $k$ are shown in Fig. \ref{fig:EDFkk}. We observe that there are only two
plateaus for the power, indicating that the spectrum is nearly scale invariant only as $k \ll 1$ and $H_{last} > k > 1$. In particular
the middle stage of acceleration does not produce an intermediate plateau in  Fig. \ref{fig:EDFkk} because the modes with
$k$ in the range of the conformal Hubble parameter during that period suffer a meltdown before and after. 
Indeed, a finely resolved
spectrum in the right panel shows the `ringing' characteristic for the modes in the radiation regime. The intermediate epoch of acceleration
simply can't turn those modes into a `permafrost'. This explicitly confirms our claim earlier that the radiation eras must be shorter than the
epochs of accelerated expansion. 
\begin{figure}[htb]
    \centering
    \includegraphics[scale=0.9]{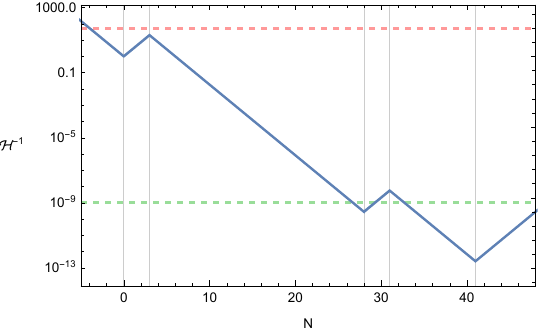}
    \caption{Inverse conformal horizon $1/{\cal H} = a/a'$ for a rollercoaster with two intermediate stages of accelerated expansion. The dashed
    lines depict the modes of the two wavelengths $1/k$, a long one (red) which freezes out early on and remains superhorizon until very late
    into final stage of radiation, and a shorter one (green) which thaws during an intermediate era of radiation and refreezes again in the second
    rollercoaster acceleration epoch.}
    \label{fig:realinvh}
\end{figure}

Our other example with three stages of acceleration now involves long stages of rollercoaster epochs of accelerated expansion, each lasting 
25 efolds, separated from each other and from the stage of preinflation by short radiation eras, each lasting 3 efolds. We take this case to
illustrate the improvement of the shape of the power spectrum when radiation epochs are shorter. This kind of improvement will occur 
generically in this limit, and we only focus on the specific example because of the computational simplifications that follow. We also note
that with our selection of duration of accelerating and radiation epochs, the CMB anisotropies can be easily matched by the perturbations
generated in this example of rollercoaster cosmology. 

\begin{figure}[htb]
    \centering
    \includegraphics[width=0.49\textwidth]{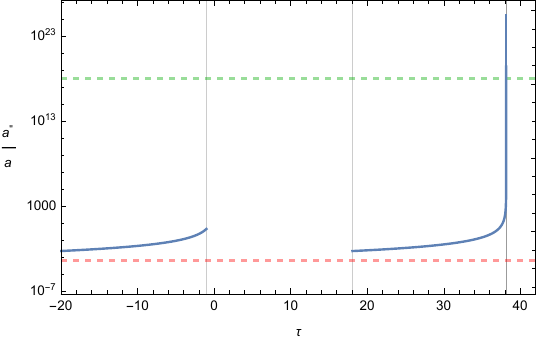}
        \includegraphics[width=0.49\textwidth]{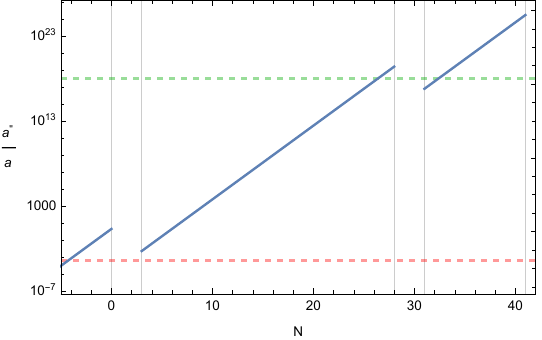}
    \caption{The perturbation potential $a''/a$ as a function of conformal time $\tau$, left panel, linear units, 
    and efolds $N$, right panel log-log units. Because of the exponentially sharp rise of the 
    potential with $\tau$ the linear plot in $\tau$ cannot discern the features of the
    final potential barrier.  
    The potential is seen in the log-log plot clearly. The dashed
    lines depict the same modes as in Fig. \ref{fig:realinvh}.}
    \label{fig:RealFkk}
\end{figure}

The numerical analysis in this case is straightforward. To explain results and develop intuition, we give the evolution of the inverse horizon 
in Fig. \ref{fig:realinvh}. The shortness of the intermediate radiation eras and the long duration of the accelerating stages is depicted by short
rises to the peaks and long downward slopes, respectively. As consequence, a mode which becomes superhorizon during an era stays 
superhorizon until very late into the terminal radiation epoch (or whatever supplants it far to the right). Those modes will preserve the power
coded by $F(k)$ determined at horizon crossing for a very long time. The review of the perturbation potential $a''/a$ adds to this conclusion,
as can be seen in Fig. \ref{fig:RealFkk}.
We see that very long wavelength modes, which at the far left is in the classically allowed region, 
eventually enter the barrier and for the most part remain in the classically forbidden regime until late into the final radiation era.
Even though the potential barriers vanish during intermediate radiation epochs, the long wavelength modes are still superhorizon and remain frozen.
Some modes reenter the classically allowed regions, and their wavelength simultaneously drops below $1/{\cal H}$.
These modes reexit again, and due to the phase shifts experience the mode ringing we have seen previously, manifested as  bumps of the power form factor $F(k)$. 
\begin{figure}[thb]
    \centering
    \includegraphics[width=0.49\textwidth]{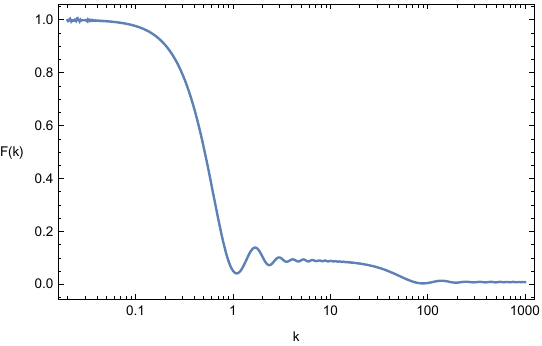}
        \includegraphics[width=0.49\textwidth]{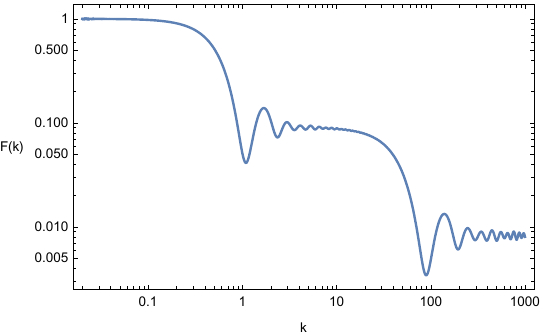}
    \caption{Power as a function of momentum; left panel, linear-log units, right panel log-log units. The power bumps are quite small in 
    overall amplitude and are best viewed in the log-log plot. {To display the results, for numerical issues, the efold duration of each stage has been rescaled by a factor of $5$.}}
    \label{fig:RealFkk1}
\end{figure}

Overall, we expect -- due to the fact that we have two long intermediate epochs of acceleration separated by short eras of 
radiation that $F(k)$ as a function of $k$ will feature three plateaus of nearly constant values: a long and very flat one for the 
preinflation era, and shorter and slightly wiggly ones for the intermediate stages of acceleration before the final era of radiation. The direct 
integration of the wavefunction confirms this. We show the results in Fig. \ref{fig:RealFkk1}. The change in power between the two
intermediate acceleration stages seem relatively small since the intermediate radiation era separating them is short; still, the log-log plot
shows that $F$ changes by over an order of magnitude -- as it must, since $H \sim 1/a^2$ during a radiation epoch, and we have
$\sim 2$ efolds for of radiation, which corresponds to expansion by a factor of $\sim$ 1000 -- 10000. Nevertheless, since the finite eras
of cosmic acceleration are longer, the bumps damp quickly and the spectrum during those eras is approximately scale invariant. It thus 
indeed seems possible that a rollercoaster with at least one long era of finite acceleration, followed by a short radiation era and additional
stages of acceleration and radiation, adding up to $\la 50$ efolds of expansion, can match the observed CMB sky.

An important feature of the spectrum shown in Fig.  \ref{fig:RealFkk1} is the stairway-like structure of the power function $F(k)$. 
We see that the function $F(k)$ -- which is a very accurate approximation of the primordial tensor perturbation spectrum -- changes 
in step-like fashion from one stage to another. This means that at short scales the tensor primordial spectrum may be very suppressed relative 
to what it is at large scales. This is a distinct signature of the rollercoaster evolution which may be used to distinguish it from, e.g. inflation.
If in the future a primordial tensor signal were observed at the largest scales on the sky, it might not indicate that a long smooth inflation
took place in the past. To confirm this we would need measurements at shorter scales. If the signal at those shorter scales disappeared, this might indicate that rollercoaster evolution took place instead. 

One might be tempted to use this observation to explain the elusiveness of the tensor signal so far. What if the CMB sky which we 
observe is a composite feature of evolution involving some breaks in cosmic acceleration? This might suppress the tensors at all but the 
longest scales on the sky, where the observations are impeded with practical issues. However this does not seem likely to us, because
no such dramatic suppressions are seen in the scalar spectrum. If the power function $F(k)$ really had a bump at a comoving momentum $k$ in the 
observable range, to hide such a bump from the scalar perturbations would require an extremely contrived form of the scalar potential and the function $r$ which we simply cannot foresee at this time. However, bumps affecting both the scalar and tensor sectors at sub-CMB scales remain perfectly viable, in our view.

\section{Discussion}

We have demonstrated in this article that cosmological problems can be addressed with the tools from an inflationist's arsenal, but deployed differently. Instead of one long smooth stage of inflation, it is possible to get the observed universe
from a sequence of bursts of cosmic acceleration, which are however significantly interrupted by stages of decelerated evolution
in between. As long as the stage corresponding to the epoch when the CMB anisotropies are generated is about 
${\cal O}(10) - {\cal O}(15)$ efolds long, and the accelerated eras after that build up additional 40 -- 50 efolds, the resulting
late universe will resemble our own.

The rollercoaster evolution leaves signatures which may be observably different from long smooth inflation, and 
allows for a greater diversity of early universe relics to survive to the present time. First off, due to the differences in the background evolution,
the background curvature $\Omega_{\tt k}$ is not getting uniformly diluted during rollercoaster evolution. 
This means that at the end of last bust of rollercoaster acceleration, while small it may be considerably bigger than at the end of long smooth inflation. This increased `survival' of the
primordial contribution breaks degeneracy between $\Omega_{\tt k}$ and $\delta \rho/\rho|_{\tt S}$, since not all the curvature-like 
behavior would be induced by a local overdensity \cite{Knox:2005hx,Kleban:2012ph,Efstathiou:2020wem}. Hence it makes sense to look
for signatures of $\Omega_{\tt k}$ larger than the scalar density contrast at largest scales. We stress however that not observing a larger $\Omega_{\tt k}$ would not automatically rule out rollercoaster due to the sensitivity of $\Omega_{\tt k}$ on the previous evolution.

The interruptions of the evolution also change how the model predictions should be fitted to the data. Instead of
what happens in long smooth inflation, the total amount of efolds comes from a combination of many different epochs, each with its own internal efold clock
and internal dynamics. The events which determine observables are timed by an efold clock intrinsic to a specific epoch, which controls
slow roll during that epoch: the overall timing still needs to be 50 -- 60 efolds before rollercoaster
ends, but this number is simply a clock reset, which has nothing to do with local conditions when, e.g., CMB 
anisotropies are generated, due to the locality of the potential in field space, and therefore locality in time. 
This changes the value of ${\cal N}$ at the CMB pivot
point, where the CMB observations are fitted to the theory. This can significantly open up the theory space, loosening the bounds. 
In particular some models that seem excluded at ${\cal N} = 60$ fit very well as shorter stages with ${\cal N} = 30$.
Specifically, the flattened power law models, such as those which arise in monodromy constructions \cite{eva1,eva2,evil,london,pisky}, fit the current data really well, and remain within reach of future searches. 

Further, due to the interruptions
by decelerating stages and the more complex structure required to drive the rollercoaster, the spectrum of perturbations
at short scales will be very different from long smooth inflation. This might be probed with future gravity wave searches.
In the explicit examples we have studied, where decelerated expansion comes from temporary radiation domination in between
consecutive stages of acceleration, the tensor power is significantly reduced relative to long inflation, implying that future searches
for primordial gravity waves at sub-CMB scales might come up empty handed. Conversely, this might make the proposal falsifiable: an 
observation of gravity waves at different scales with comparable power would put paid to our suggestion. There may also be differences in the scalar power spectrum. For example, we noted that in shortened stages of acceleration the running of the spectral index would
be enhanced. This could affect the evolution of the CMB structures, as pursued in for example \cite{sarblan}, although it seems that the data has tightened on this issue. Nevertheless, some purported anomalies might still lurk there.

However there is some
model dependence here that needs to be underscored. If the epochs separating stages of acceleration are not dominated by
radiation, but by something else, the evolution of power could be different. Some examples might be matter domination \cite{Peter:1994dx},
or even more dramatically controlled bounces such as in the proposal of \cite{Graham:2017hfr}, that might even generate 
a universe resembling the picture in \cite{Ijjas:2019pyf}. If the intermediate stages are bounces instead of radiation, the 
interim thawing of
modes which we have seen in the case of radiation might never occur. The modes during bounces could continue to freeze out,
albeit differently than during accelerating eras. The bottomline is that the spectrum would be different than in long smooth inflation,
and to find the differences one does need to provide a detailed model. 

Finally, the abundance of reheating products from previous accelerated stages does not get completely wiped out. The dilution comes only from 
the accelerated stages after the population of particles, or for that matter, even primordial black holes, is born. 
If heavy enough, such objects, although more rare than the visible ones, may be dark matter. As an example consider
primordial black holes. Imagine that during the next-to-last stage of inflation, the dynamics was driven by some field with a  potential like that of new inflation \cite{Linde:1981mu,Albrecht:1982wi}, with the field on a hilltop. During such evolution,
the density perturbations created at the end would be large, easily leading to $\delta \rho/\rho|_{\tt S} \sim {\cal O}(1)$,
at scales much shorter than the scales of CMB anisotropies. These perturbations give the right conditions for the formation
of black holes of the size of cosmological horizon at that time \cite{Carr:1974nx}.  It is interesting to estimate their abundance 
now. 

If these black holes are to survive until the present time, they have to be cold enough to suppress Hawking radiation.
This means their mass has to be at least $m_{BH} \sim 10^{12} {\rm kg} \sim 10^{30} {\rm GeV}$. However black holes
so small are constrained more strongly, and it seems they cannot be all of dark matter 
\cite{watfries,Ballesteros:2019exr,Green:2020jor}. 
Yet the bounds weaken sharply with the increase of mass, and already at $m_{BH} \sim 10^{33} {\rm GeV}$
appear negligible. The horizon size of black holes with $10^{33} {\rm GeV}$ mass is $r_{BH} \sim m_{BH}/M_{Pl}^2 \sim 1/{\rm TeV}$,
and thus their temperature is $T_{BH} \sim {\rm TeV}$. If these black holes were formed cosmologically, by a collapse of a horizon-size
overdensity \cite{Carr:1974nx}, the cosmological horizon at that time had to be $1/H_* \ga r_{BH}$, and so $H_*  \la T_{BH}$. 
As the formation rate estimates indicate \cite{watfries,Green:2020jor}, the black hole formation 
probability should be below unity, which means that the holes would not form inside every Hubble patch of size $1/H_*$. This should ensure that the
black holes do not overclose the universe when they form. 

To 
make a crude estimate of the resulting late time abundance, we will normalize the black hole population by requiring 
that the nearest neighbors are initially about a thousand horizons apart, to have a chance to describe their evolution
perturbatively. This amounts to one black hole per one billion Hubble volumes. This gives us their energy density at the time $t_*$ of their formation to be 
$\rho_{BH}(t_*) \sim  10^{-9} \, m_{BH} H_*^3 \sim 10^{-9} M_{Pl}^2 H_*^2$. Due to their mass, these black holes will 
evolve as nonrelativistic matter, and their energy density will dilute as $1/a^3$ as time goes on. So imagine that 
after they are formed, there is a stage of about two efolds of radiation, and five-six more efolds of accelerated expansion, before
the final reheating. The energy density in these black holes will dilute, but only by a factor of about $a^3 \sim 10^{11} - 10^{12}$,
becoming about $\rho_{BH}(t_{end}) \sim 10^{-21} M_{Pl}^2 H^2_{end} - 10^{-20} M_{Pl}^2 H^2_{end}$ where $H_{end}$ is the Hubble parameter at the end of the rollercoaster. 
It is about $\sim H_*/100$ due to the two efolds of radiation separating the two last stages of accelerated expansion. 
Assuming radiation domination from that point on to the time when the temperature of the universe is about eV, at radiation-matter 
equality, we find that the fraction of the energy density in black holes at radiation-matter equality grows to
\be
\frac{\rho_{BH}(t_{eq})}{\rho(t_{eq})} \sim \frac{\rho_{BH}(t_{end})}{\rho(t_{end})} \frac{T_{end}}{T_{{eq}}} \, ,
\label{rhos}
\ee
where the factor ${T_{end}}/{T_{eq}}$ accounts for the fact that black hole energy density scales as $T^3$ versus $T^4$ for radiation.
Here $T_{end} \sim \sqrt{M_{Pl} H_{end}} $ is the reheating temperature at the end of the rollercoaster, and
with our numbers it is $T_{end}\sim 10^{11} {\rm GeV} \sim 10^{20} {\rm eV}$. Substituting this into (\ref{rhos}), 
somewhat unexpectedly, we find that at radiation-matter equality $\rho_{BH}/\rho$ could be ${\cal O}(1)$, 
\be
\frac{\rho_{BH}(t_{eq})}{\rho(t_{eq})} \sim 0.1 - 1 \, ,
\label{rhosnos}
\ee
which suggests that all of dark matter just might be primordial black holes
left over after the early universe rollercoaster ride, although we admit that our estimate is  
somewhat crude. However the coincidence is interesting, and we wonder if the argument can be refined; we stress
that similar ideas have pursued in the context of double inflation previously, for example in \cite{yanagida}. 
Note that there is some wiggle room in this argument; for example if black holes are more massive or denser, the extra contribution to energy density could be diluted a little
more with the longer final stage of acceleration in the rollercoaster. This can also account for uncertainties 
due to the mass variation of the black hole population, induced by matter accretion.
However 
we cannot arbitrarily increase the black hole mass in this argument,
since that would require decreasing the Hubble parameter at the end of the rollercoaster evolution, which in turn 
would require more expansion afterwards to dilute energy density. This indicates that there may be a `sweet spot' 
in the range of black hole masses that could be produced in rollercoaster cosmology. It remains to be seen if this 
might be developed into an analogue of the
old wimp miracle in the rollercoaster context.

\section{Summary}

We have presented the framework for rollercoaster cosmology and showed that it can fit the current observations. Rollercoaster evolution solves the 
standard cosmological problems by smoothing out the background geometry just like inflation. 
This gives opportunities for much model building and exploration of cosmological observables and predictions. 
The key insight which opens the door to this approach to early universe cosmology is that the common belief that 
there was a long smooth stage of inflation which lasted 50 -- 60 efolds is really just an {\it extrapolation} from the observational viewpoint. It fits the data where we can look, but we do not have 
any direct evidence that inflation occurred in one go beyond the scales which we explore with 
the CMB anisotropies, give or take. It may also lead to different
approaches in model building which might be easier to reconcile with UV completions, since the shorter stages of acceleration may be 
less sensitive to Planck scale physics and more resistant to perils of quantum gravity.

In addition, we have seen that tensor perturbations -- i.e. primordial gravity waves -- are a model-independent
probe of the cosmic evolution. Once we declare gravity to be described by GR, the tensor power will be an
excellent indicator of the background evolution, staying roughly constant during epochs of acceleration
when $\dot H \simeq 0$, 
and quickly decreasing when $\dot H$ turns on. Of course, this follows directly from equivalence principle. If such signals can be detected, we would find a useful tool
for cosmic archeology. 

In a way our proposal is akin to the process of charting the Great Desert of particle physics, where the extrapolation of low energy data
under the assumption that nothing intervenes for decades of scales yields unification at the GUT scale. As we know
now, this may not be correct. An example is provided in the theories with large extra dimensions \cite{ADD}, 
which while currently constrained by the data offer a different perspective. We suggest that a similar situation
might occur in cosmology, where the `time desert' of long smooth inflation might in fact be interspersed with
multiple oases of different physics. The rollercoaster cosmology 
approach provides a different way to scan through these scales, and populate them with signatures which may be observably different from 
long smooth inflation. 

In light of this, it therefore appears that sometimes adopting a mathematician's logical stance may be essential 
and indispensable in interpreting cosmological  data. Have patience, the Universe is not finished with us yet. 

\vskip.5cm

{\bf Acknowledgments}: 
We would like to thank D.E. Kaplan, S. Rajendran, J. Terning and A. Westphal for useful comments and discussions. NK is supported in part by the DOE Grant DE-SC0009999.

\end{document}